\newcommand{\mode}{1}

\ifnum\mode=0 
\documentclass[AMA,STIX1COL]{WileyNJD-v2}
\articletype{Article Type}%

\received{26 April 2016}
\revised{6 June 2016}
\accepted{6 June 2016}

\setcitestyle{numbers,square} 

\else 
\documentclass[11pt,a4paper]{paper}

\usepackage[top=1in, bottom=1in, left=1.25in, right=1.25in]{geometry}

\usepackage{graphicx}
\usepackage{enumerate}
\usepackage[font=small]{caption}
\usepackage{amsmath}
\usepackage{amsthm}
\usepackage{amsfonts}
\usepackage{url}
\usepackage{algorithm, algorithmicx, algpseudocode}
\usepackage{authblk}
\usepackage{natbib}
\usepackage[colorlinks=true,citecolor=blue,linkcolor=]{hyperref}
\setcitestyle{numbers,square} 

\fi 

\newcommand{\vz}{\ensuremath{\mathbf{0}}}
\newcommand{\vx}{\ensuremath{\mathbf{x}}}
\newcommand{\vL}{\ensuremath{\mathbf{L}}}
\newcommand{\vn}{\ensuremath{\mathbf{n}}}
\newcommand{\vN}{\ensuremath{\mathbf{N}}}
\newcommand{\vX}{{\ensuremath{\mathbf{T}}}}
\newcommand{\vuv}{\ensuremath{\mathbf{uv}}}
\newcommand{\vY}{{\ensuremath{\mathbf{S}}}}
\newcommand{\vu}{\ensuremath{\mathbf{u}}}

\newcommand{\vF}{\ensuremath{\mathbf{X}}}
\newcommand{\reals}{\ensuremath{\mathbb{R}}}

\newcommand{\surface}{\ensuremath{\mathcal{S}}}
\newcommand{\curve}{\ensuremath{\mathcal{C}}}
\newcommand{\norm}[1]{\left\lVert#1\right\rVert}

\newcommand{\ie}{\textit{i}.\textit{e}. }
\newcommand{\eg}{\textit{e}.\textit{g}. }
\newcommand\rsec[1]{Sec. \ref{#1}}
\newcommand\reqt[1]{Eq. \ref{#1}}
\newcommand\rfig[1]{Fig. \ref{#1}}

\begin{document}

\title{Quasi-structured quadrilateral meshing in Gmsh -- \\a robust pipeline for complex CAD models}

\ifnum\mode=0  

\author[1]{Maxence Reberol*}

\author[1]{Christos Georgiadis}

\author[1]{Jean-Fran{\c c}ois Remacle}


\authormark{Reberol M. \textsc{et al}}

\address[1]{\orgname{Universit\'e catholique de Louvain}, \country{Belgium}}

\corres{*Maxence Reberol,
     Avenue Georges Lemaitre 4, Louvain-la-Neuve, Belgium.
\email{maxence.reberol@uclouvain.be}}

\else 

\date{}
\author[1]{Maxence Reberol}
\author[1]{Christos Georgiadis}
\author[1]{Jean-Fran{\c c}ois Remacle}
\affil[1]{Universit\'e catholique de Louvain, Belgium}

\fi 


\ifnum\mode=0  
\abstract[Summary]{
\else 

\maketitle

\abstract{
\fi 
    We propose an end-to-end pipeline to robustly generate high-quality
    quadrilateral meshes for complex CAD models.  An initial quad-dominant
    mesh is generated with frontal point insertion
    guided by a locally integrable cross field
    and a scalar size map adapted to the small CAD features.
    After triangle combination and midpoint-subdivision into an all-quadrilateral mesh, the topology of
    the mesh is modified to reduce the number of irregular vertices.
    The idea is to preserve the irregular vertices matching
    cross-field singularities and to eliminate the others.
    The topological modifications are either local
    and based on disk quadrangulations,
    or more global with the remeshing of
    patches of quads according to predefined patterns.
    Validity of the quad mesh is guaranteed by monitoring
    element quality during all operations and reverting the changes when necessary.
    Advantages of our approach include robustness, strict respect of the CAD features
    and support for user-prescribed size constraints.
    The quad mesher, which is available in Gmsh, is validated and illustrated on
    two datasets of CAD models.
}

\ifnum\mode=0  

\fi 


\ifnum\mode=0  
\maketitle
\fi 


\section{Introduction}

There are essentially two main scientific communities that have been
historically interested in generating quadrilateral meshes: the engineering
analysis community, as these meshes are good geometrical support for the finite
element and finite volume methods, and the computer graphics community 
which, besides application in numerical simulations, has good use for them 
in surface modeling (\eg subdivision surface) and texturing.

The engineering analysis community has been working on algorithms for
generating quadrilateral meshes for several decades, leading to two main
categories of industrial-grade techniques. The first one is to manually
decompose the CAD model in quadrilateral patches, with potentially some
semi-automatic assistance.  The patches are then filled with quads while
considering user element sizing specifications,\eg anisotropic quads with
geometric progression from the boundary layers. This approach is very time-consuming
and is mainly used
for demanding numerical simulations such as CFD. On the other end, various
fully automatic approaches have been developed for generating unstructured
quadrilateral meshes, such as paving \cite{blacker1991} or Blossom-Quad
\cite{blossom} with frontal Delaunay point insertion \cite{delquad}. Their
disadvantages are the potentially high number of irregular vertices, the
non-regular alignment of edges in the mesh (affecting smoothness of gradients
in simulations) and the lack of user control. Both approaches
are now available in most mesh generation packages, including Gmsh \cite{gmsh2009}. 

In the last decade, the computer graphics research community has thoroughly
developed and explored cross-field based techniques for quadrilateral meshing,
with a focus on the automatic generation of high-quality block-structured
quadrilateral meshes, or equivalently coarse \emph{quadrilateral layouts} which
partition the object surfaces into conforming quadrilateral patches.
Finite element practitioners quickly became interested in these
ideas because of their potential to replace semi-automatic block-structured
mesh generators by fully automatic ones. A fruitful line of research revolves
around the construction of a global parametrization from which we can extract
an all-quadrilateral mesh. The global parametrization is a discontinuous mapping
that maps a 2D grid from the parameter domain to the model surface, with
discontinuities to allow for the existence of irregular vertices 
(\rsec{ss:related_work} for more details). The work
presented in the current paper started from an attempt to apply such a global
parametrization technique in the context of CAD meshing in Gmsh, as explained
later (\rsec{ss:motivation}).

The quasi-structured quadrilateral meshing pipeline that we propose exploits
the strength of both worlds. We rely on the robustness of automatic
unstructured quad meshing techniques, and we use the topological and geometrical
information from cross fields to transform unstructured meshes into
high-quality ones, with few but well located irregular vertices.
Our approach has many advantages which are important for CAD meshing: it always produces
valid quadrilateral meshes, it strictly respects all CAD features and
it supports non-uniform element size maps.
The new quad mesher is integrated in Gmsh (version 5 and higher) \cite{gmsh2009}
and its implementation is open-source.

\begin{figure}
    \begin{center}
        \includegraphics[width=\textwidth]{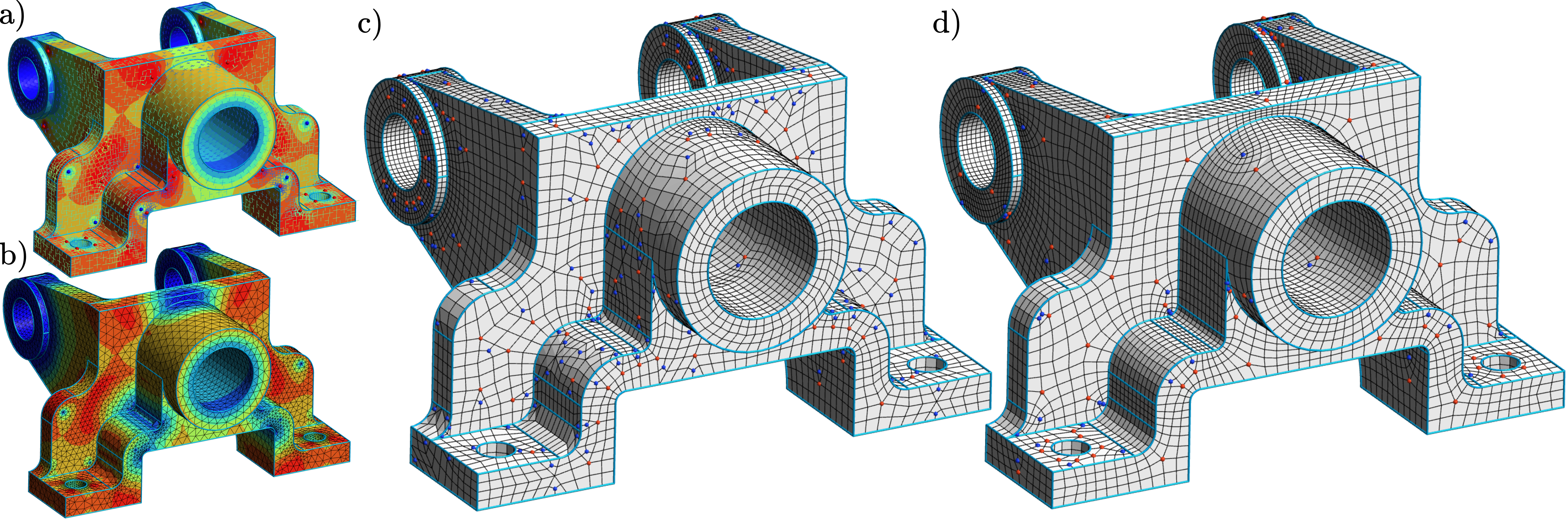}
        \caption{The quasi-structured quad meshing pipeline applied on the model
            \emph{M3} \cite{mambo} (111 CAD corners, 169 curves, 60 faces).
            The cross field and its associated conformal scaling is computed
            on an initial triangulation (a). The size map (b) combines the 
            cross field scaling and is adapted to the small CAD features.
            The initial unstructured all-quadrilateral mesh (c) is computed with
            a frontal approach, and contains $3429/54982$ irregular vertices.
            After cavity remeshing, the final quasi-structured mesh (d) contains
            $606/56743$ irregular vertices. $78$ of them match the cross field singularities
            (a) and the others allow the mesh size transitions (b).
        }
        \label{fig:M3}
\end{center}
\end{figure}

\subsection{Overview}

This paper proposes a robust end-to-end pipeline to build quasi-structured
quadrilateral meshes for arbitrary CAD models. Our only requirement of the
input CAD model is that it is valid in the sense that it is possible to generate a
triangulated mesh with standard techniques. This includes non-manifold
surfaces, which typically appear when the CAD models contains multiple volumes.
By quasi-structured quadrilateral mesh, we mean that the number of irregular
vertices is small compared to the usual unstructured techniques, and close or equal
to the number of irregular vertices obtained with cross-field based
parametrization approaches when the target mesh size is smaller than the CAD
features (\eg \rfig{fig:gparam_vs_qqs}). Our meshing pipeline, illustrated
on \rfig{fig:M3}, is:

\begin{enumerate}[\hspace{1cm}Step 1:]
    \item Compute a scaled cross field on each CAD face, using an initial triangulation.
        The cross directions are computed with a fast multilevel heat diffusion approach
        and a conformal scaling is obtained by solving a linear system, see \rsec{sec:cross}.
    \item Compute a global metric field (combined direction field and scalar size map) from
        the prescribed sizing constraints, the small CAD features and the conformal scaling associated to
        the cross field, see \rsec{sec:sizemap_and_curve}.
    \item Generate the meshes of the CAD curves considering the size
        map and global topological constraints, see \rsec{ss:curve}.
    \item Generate the quasi-structured quadrilateral meshes of the simple CAD faces which match
        predefined patterns, see \rsec{sss:simple_face}.
    \item Generate unstructured quadrilateral meshes on the remaining CAD faces, using a frontal approach 
        guided by the cross field scaled by the size map, see \rsec{sec:unstructured}.
    \item Modify the quadrilateral mesh topology to make it quasi-structured,
        by removing defects with local disk-quadrangulation operations (\rsec{sec:disk_quad})
        and by remeshing convex cavities to match the cross field singularities (\rsec{sec:cavity}).
\end{enumerate}

With this pipeline the ultimate focus is on robustness, which has
precedence over all other considerations. The unstructured quad mesh
is geometrically and topologically valid by construction (quad-dominant mesh
subdivided into full-quad). For all the remeshing
operations, we always verify that the topology and the geometry remain
valid, and we reject the ones which invalidate or degrade the element qualities.

The scaled cross field guides the unstructured point insertion and its
singularities are used to constrain the cavity remeshing. All the steps
produce valid outputs even if the cross field is not accurate or is topologically invalid.
The key idea to robustly deal with complex CAD models is to use the
cross field as an auxiliary tool without entirely relying upon it, 
since it may and usually contains many defects 
(does not take small CAD features or quantization into account, may have wrong
singularities, may have insufficient accuracy).

Figure \ref{fig:gparam_vs_qqs}.2. illustrates our pipeline for a single
surface. In practice, this surface should be seen as a particular CAD face of a
larger model, so its 1D boundary mesh cannot be modified without affecting its
neighbors. This compatibility constraint between the meshes of adjacent CAD
faces is one of the major sources of difficulty when dealing with complex CAD 
models. We can look at a more realistic CAD model on \rfig{fig:M3}, where the
CAD faces are meshed independently. After topological improvement (\rfig{fig:M3}.d.),
there are still irregular vertices that do not match singularities from
the cross field. These are due to the curve quantization which follows the
size map (\rfig{fig:M3}.b.), and which is necessary to take small CAD features
into account. For more detailed discussions and applications to complex
models, see \rsec{sec:results}.

Because we present an end-to-end mesher, a lot of ground is covered and it
is not practical to describe everything in details. We try to provide
extensive references to the literature so the reader
can refer to them for further details on specific parts of the pipeline.

\subsection{Related work}
\label{ss:related_work}

\paragraph{Unstructured quadrilateral meshing}
A classic approach to generate unstructured quadrilateral meshes is to start
from the boundary and to progressively insert points in the surface. 
The \emph{paving} algorithm \cite{blacker1991}
directly generates quadrilateral in an advancing-front fashion.
Indirect methods first generate a triangulation with well-placed vertices then merge pairs
of triangles into quads \cite{leeNewSchemeGeneration1994,frey}. 
There are many ways on how to choose the merged triangles. In Q-Morph
\cite{owenQMorphIndirectApproach1999}, triangles are transformed in
quadrilaterals with an advancing-front algorithm. BlossomQuad \cite{blossom}
computes a global perfect matching to optimally pair triangles,
which works especially well with right-angled triangles generated
by \cite{delquad}.

These methods generally suffer from degraded mesh quality and high vertex
irregularity, especially at front collisions.  But on the other hand, they are
highly robust.  Even when a pure quad mesh cannot be obtained directly, it's
always possible to perform a midpoint subdivision, also called Catmull-Clark
subdivision, to obtain an all-quadrilateral mesh by splitting the triangles into
three quads and the initial quads into four sub-quads.

\paragraph{Cross fields}

In the last decade, cross fields \cite{hertzmann2000,Ray2006,Palacios2007} have
become very popular for quadrilateral meshing.  Crosses can be seen as
infinitesimal quads and they have topological structures which match
quad mesh topology, such as cross singularities which have one-to-one
correspondence with quad mesh irregular vertices \cite{bunin2007,Ray2008}.

Cross field can be computed by minimizing a smoothness energy, either by
directly optimizing the angles \cite{bommes2009} or by smoothing a vector
representation with a way to ensure that the variables stay crosses, such as
Ginzburg-Landau penalization \cite{beaufort2017} or with iterative projection
\cite{viertel2019}. In this paper, we use the MBO method \cite{viertel2019} where
we alternate diffusion and projection, with varying time stepping \cite{palmer2020}
and additional optimizations (\rsec{ss:heat}).

It is important to note that cross fields are not \emph{integrable} by default.
On a cut domain, a cross field can be seen as two orthogonal unit vector fields.
To have integrable vector fields, it is necessary to break
the orthogonality \cite{sageman2019} or to scale the amplitudes. 
Ray N. et al. \cite{Ray2006} introduce a correction step to make the two vector fields
curl-free. Another approach it to solve a linear problem to compute a global conformal
scaling from the cross field singularities \cite{bunin2007,ben2008}, which can be applied
to both vector fields. We compute a similar conformal scaling (\rsec{ss:scaling})
by canceling a Lie bracket, without using the cross field singularities.

There are various ways to exploit the cross field information for quad meshing.
It can be used as a directional guide for frontal point insertion, as in \cite{delquad}
and in our unstructured initialization (\rsec{sec:unstructured}). 
A popular alternative is to directly build the quadrilateral layout by
tracing separatices in the vector fields, \eg \cite{campen2012,myles2014,pietroni2016}.
Another common approach is to \emph{integrate} the cross-field by computing a parametrization.

\paragraph{Cross field based parametrization} 
A parametrization associated to a cross field is usually made
of two scalar fields $u,v$ such that their gradients are aligned
with the cross field directions. Such parametrization is a mapping
from the 2D parameter space to the 3D surface.  The mapping discontinuities,
which are cuts on the model surface, allow for the existence of irregular
vertices on the quadrilateral mesh. QuadCover \cite{quadcover}
introduced globally coherent parametrization to \emph{integrate} a cross field,
with integer grid-preserving constraints on a cut-graph to ensure global
coherency. This idea have been highly refined and improved in the form of
integer grid mapping \cite{bommes2009,bommes2013,campen2015,qex}.
With these approaches, the irregular vertices in the final quad mesh strictly
match the cross field singularities. Alternatively, the parametrization 
formulation can be relaxed, either by
computing local parametrizations \cite{jakob2015} or a loosely constrained 
global
parametrization which relies on periodic functions \cite{Ray2006}. In both
cases, there are additional irregular vertices in the final quad mesh
introduced by the relaxation.
Our experience with a global parametrization attempt is detailed in
\rsec{ss:motivation}.

\paragraph{Topological meshing and modifications}
Unstructured quadrilateral meshing methods do not naturally offer
the regularity and structure obtained with global parametrization methods.
Is it possible however to reduce the number of irregular nodes and improve 
quality by performing topological modifications on an unstructured quad mesh.
Due to topological constraints in quadrilateral meshes (\rsec{sec:topo_quad}),
there is no simple local operation such as the edge-split or edge-collapse,
but there are still many possibilities, some global such as chord collapsing \cite{Daniels2008}
and other more local \cite{sanchez} but which are restricted to specific vertex-valence configurations. 

In this work, we are particularly interested in operations that modify the
content of a patch of quads inside the mesh, called cavity, without changing
its boundary.  For triangular and pentagonal cavities, Bunin G. \cite{bunin2008} has
shown that we can find a topological quad mesh replacement with a single
irregular vertex by solving a linear system. This local clean-up
is a fundamental piece of the Jaal quad mesher \cite{jaal}, with
a later extension to more generic replacement patterns \cite{verma2015}.

This problem of remeshing a cavity is equivalent to the problem of
topologically finding a quadrilateral mesh for a given polyline boundary.
It has been studied in the computer graphics context \cite{yasseen2013},
with a generic integer linear formulation for generic $N$-sided patches
\cite{takayama2014}. The local remeshing operator has been used
in interactive software to help users improve their quad meshes \cite{marcias2015}.

\subsection{Topology of quadrilateral meshes}
\label{sec:topo_quad}

In a polygonal mesh, the Euler-Poincar\'e formula $\chi = n - n_e + n_f$ provides a
relationship between the $n$ vertices, the $n_e$ edges, the $n_f$ facets and 
the Euler-Poincar\'e characteristic $\chi$  of the associated surface, which is
given by the relation $\chi = 2 - 2g - b$ with $b$ the number of holes and $g$
the number of handles.

In a quadrilateral mesh, we can follow \cite{beaufort2017} to derive the relationship
$$4 n_f = \sum_k 4 (n - n_b - n_k) + 2 (n_b - m_k) + (4-k) n_k + (2-k) m_k$$
between the number of vertices of various indices. $m_k$ is the number of boundary
vertices with valence $2-k$ and $n_k$ the number of interior vertices with valence $4-k$.
When $k\neq 0$, the vertices are said to be irregular of index $k$.
By combining this relation with the Euler formula, we obtain \cite{beaufort2017} the relation
\eqref{eq:euler_quad} between the number of irregular vertices on the boundary and on the interior
in a quadrilateral mesh:
\begin{equation}
    \label{eq:euler_quad}
    \chi = \sum_k {k\over 4} (n_k+m_k)
\end{equation}

This formula has some interesting topological implications: regular vertices do not count,
pairs of irregular vertices of opposite indices cancel themselves (\eg valence
three ($k=1$) and valence five ($k=-1$)), surfaces with $\chi=0$ (topological
ring) are the only ones that admit fully regular quadrilateral meshes (only valence four inside and
valence two on the boundary).

We will see later in this paper that it is possible to generate
quadrilateral meshes with interior irregular vertices of indices $+1$ and $-1$ only,
corresponding respectively to quadrilateral valences of $3$ and $5$.  
This restriction to irregular vertices of low index is usually better
to generate high-quality quads, as irregular vertices are associated with 
geometric distortions.
In this paper, we thus restrict ourselves to irregular boundary vertices of valence $1,3$ and $4$,
\ie of indices $1,-1$ and $-2$, and to interior irregular vertices of valences $3$ and $5$, \ie of
indices $+1,-1$. The relation \eqref{eq:euler_quad} between the surface Euler-Poincar\'e characteristic
and the quadrilateral mesh irregular vertices becomes:
\begin{equation}
    \label{eq:euler_quad_2}
    n_{1} - n_{-1} = 4 \chi + m_{-1} - m_{1} + 2 m_{-2} 
\end{equation}

We say that a quadrangulation of a surface with Poincar\'e characteristic $\chi$ is
\emph{minimally irregular} if its contains exactly $4|\chi|$ isolated irregular
vertices of index $\mbox{sign}(\chi)$ (counting the boundary irregular vertices).
For example, in the surface of
\rfig{fig:gparam_vs_qqs}, we have $\chi = -6$ and $m_1 = 4$ convex corners,
so we should have $n_{-1} = 28$ irregular vertices of valence five to be
\emph{minimally-irregular}.

A dipole is a
pair of irregular vertices with opposite indices $+1$ and $-1$.
Adding \emph{dipoles} to a minimally irregular quad
mesh may be necessary, for example to allow
mesh size variations. 
A mesh is called \emph{quasi-structured} if
it contains $4|\chi|$ isolated irregular vertices of index
$\mbox{sign}(\chi)$ plus a small amount of dipoles. 
We use the loosely defined adjective \emph{small} because it is {\it a
  priori} impossible to determine the minimum amount of dipoles that
are necessary for a given application.


\section{Context and motivation}
\label{ss:motivation}

The more general context around this article is our effort to build a robust
high-quality quadrilateral mesher in \emph{Gmsh} that actually \emph{works}.
What we mean is that a user should be able to produce a high quality quad mesh on any
valid CAD model by only \emph{loosely} specifying the target element sizes, 
as it is the case for triangular meshing.

\begin{figure}
    \begin{center}
        \includegraphics[width=1\textwidth]{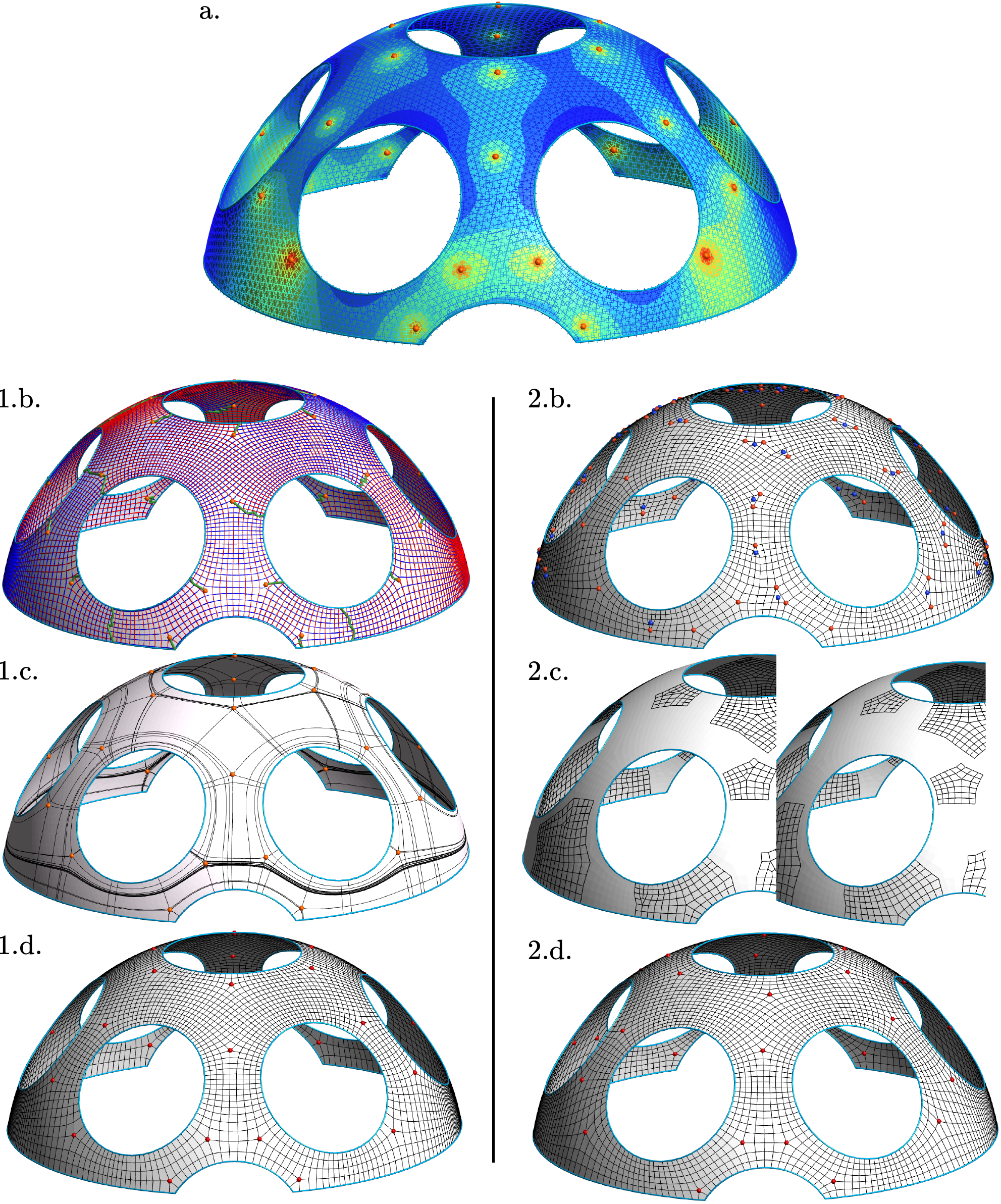}
        \caption{
            Two approaches to build a quasi-structured mesh of a surface.
            In the global parametrization approach, we build a seamless parametrization (1.b.)
            by integrating a scaled cross field (a.), then we extract a conformal quadrilateral
            layout (1.c.) and build the final quad mesh (1.d.) by simplyfing the subdivided
            layout. In the second approach (this paper), we first build a unstructured
            quadrilateral mesh (2.b.) with a frontal approach guided by the 
            scaled cross field (a.), then we apply cavity remeshing (2.c.) to eliminate
            unnecessary irregular vertices while preserving the cross field singularities.
            The final quad mesh (2.d) is very similar to the one obtained with the
            global parametrization approach and has the same number of irregular vertices.
        \label{fig:gparam_vs_qqs}}
    \end{center}
\end{figure}

\paragraph{Global parametrization} 
Our initial attempt, inspired by state-of-the-art global parametrization
techniques \cite{bommes2009,campen2015}, was to build a seamless
uv-parametrization from a cross field and then extract a quad-layout. 
Figure \ref{fig:gparam_vs_qqs}.1. illustrates the different steps for the
construction of a quadrilateral mesh with this approach.
Parametrization based techniques cut and flatten the surface of the
domain. The cuts (in green on Figure \rfig{fig:gparam_vs_qqs}.1.b.) allow to 
find a continuous branch of the cross field and to build a seamless
$uv$ parametrization (see \rfig{fig:gparam_vs_qqs}.1.b.)
by solving a linear system with an alignment
energy (gradients of $u$ and $v$ must be parallel to the cross field)
and specific grid-compatibility linear constraints on the cut graph.
To avoid the technical difficulties associated with the integer variables in the
standard mixed-integer formulation \cite{quadcover,bommes2009}, we directly
extracted the (conformal) quadrilateral layout from the iso-value separatrices
in the $uv$ parametrization (\rfig{fig:gparam_vs_qqs}.1.c.). We replaced the integer
variables and associated global quantization \cite{campen2015} by explicit
topological simplification (chord collapse) of the quad layout, since we initially
had lot of very thin quad patches as we were doing no snapping or geometric
thresholding. The resulting quadrilateral mesh (\rfig{fig:gparam_vs_qqs}.1.d.) was
of very high quality, when it worked.

Even though we were quite excited with our initial results, we failed
to generalize the approach into a robust mesher able to handle large and
complex CAD models.  It took us considerable amount of time and energy to
obtain an implementation that is still in a very beta version
state.
The amount of non-reusable code that is required for generating the quad mesh
of \rfig{fig:gparam_vs_qqs}.1.d. is quite large and only works when the cross
field topology is in perfect accordance with the topology of the surface to
mesh, which is often not the case when smoothing continuous representations of
crosses. On a given CAD face, the difficultly mainly comes from the linear system that computes
the $uv$-parametrization: it must be strictly right. If there is a small
incoherence in the grid-compatibility constraints, \eg due to wrong flagging
of singularities, there will be no solution, or a useless one.
Even more difficult is the generalization to large CAD models with hundreds of
CAD faces, potentially non-manifold at CAD curves if there are multiple volumes.
To avoid separatrices winding up around the models for thousands of turns before
coming back to their origin, we build independently the quad layouts on each CAD
face and then try to reunite them. The union of the quad layouts has many
T-junctions on the CAD curves. One potential idea is to use a global integer
quantization \cite{campen2015} to assign compatible non-zero integers to
each edge of the global quad layout (with T-junctions). However, we soon
discovered that often there is no non-degenerate solution: the constraints
can be circular, \eg $a+b=c, c = d, d = e, e = a$ leading to $b=0$.

While some issues could be fixed, we do not foresee a robust implementation of that algorithm that would
work for industrial grade models within a reasonable timescale.  Moreover, such
long codes are prone to bugs and are difficult to maintain.  Of course, the
mesh shown on \rfig{fig:gparam_vs_qqs}.1.d. is exactly the what most finite
element practitioners are willing to use. The main question became:

\vspace {.2cm}
\noindent{ \emph{Can we obtain such high quality mesh using an
    approach that is both simple and robust?}}
\vspace {.2cm}

\paragraph{Unstructured quad meshing} 
Our experience in indirect quad meshing
\cite{blossom,delquad} has been positive in term of industrial
use. Indirect quad meshers are essentially based on existing surface
meshers (very small amount of extra coding). The
transformation of a triangular mesh into a quad mesh by combination
of pairs of triangles is an extra step that is simple, robust and
efficient \cite{blossom,frey}. Indirect quad meshers are thus able to treat 
models of arbitrary complexities with a run time that is not larger
than standard triangular meshing.
The main drawback of quad meshes generated in an indirect fashion is
that they cannot be considered as block-structured or quasi-structured
because they contain an arbitrary large number of irregular vertices.

All those reasons have convinced us to move to a new approach that
would take \emph{the best of both worlds}. We have decided to use
our existing indirect meshing technology and build an \emph{initial unstructured quad
    mesh} (\rfig{fig:gparam_vs_qqs}.2.b.) that is mostly aligned with a \emph{cross
field}. Then, information provided by the cross field is used   
to modify this unstructured mesh and make it \emph{quasi-structured},
\ie with only a few irregular vertices.

Besides working even if the cross field is inaccurate, there is another big advantage:
we can use a external non-uniform size maps to locally control the mesh
size, independently of the cross field. In practice, we compute a size map
adapted the small CAD features, which has proven to be invaluable when
dealing with complex models.

\paragraph{Gmsh surface-to-volume philosophy} 

Gmsh's usual meshing pipeline is essentially "surface to volume", also called
"bottom-up". Model curves are discretized (meshed) at first. Then, model
surfaces are triangulated/quadrangulated based on a boundary discretization
that is imposed by the model curves. Finally, model volumes are tessellated.
This approach has many advantages: strict respect of CAD features, flexibility
(different meshers can be used for different CAD entities), robustness (we can 
replace with a different approach on a entity if one failed), easy parallelism, and the
automatic support for non-manifold surfaces in
models with multiple volumes. Sometimes there is no solution for the meshing of
an entity given its boundary mesh (\eg too coarse). In these cases, the
boundary mesh is modified (usually a split of some elements) and the meshing of
the entities that share this boundary is restarted, until a solution is found
or the maximal number of tries is exhausted.

We also follow this philosophy with our quad mesher to benefit from the same
advantages, but it should be noted that it has a significant drawback when
meshing a CAD face: the boundary mesh (from the boundary CAD curves) is locked
and cannot be changed. Thus, the curve quantization (\rsec{ss:curve}) is very
important. The size map, which controls the curve sampling, must be global
and smooth on the CAD curves to have coherent meshes on the adjacent CAD faces.
With our current approach, there are some irregular vertices in the final mesh that
are due to non-optimal curve quantization, and which could be eliminated with 
further work (\rsec{sec:future}).

\section{Scaled cross field computation}
\label{sec:cross}
In our pipeline, cross fields fulfill two roles: (a) they guide the
frontal point insertion in the unstructured quadrilateral mesher
(\rsec{sec:unstructured}) and (b) they have singularities (\rsec{ss:singularities})
which correspond to irregular
vertices that we want to keep during the topological modifications
that lead to the final quasi-structured mesh (\rsec{sec:cavity}).
The cross field is computed with a fast multilevel diffusion
algorithm (\rsec{ss:heat}) which alternates diffusion steps and projection
on the cross space.
Then, to improve the accuracy of the following point insertion, we scale the cross field
to make it integrable by computing its conformal scaling (\rsec{ss:scaling}), 
which can be seen as a size map (up to a constant) useful to accommodate rotations
in the cross field.

%

\subsection{Fast cross field computation with multilevel diffusion}
\label{ss:heat}

A cross field is a field defined on a surface with values in the
quotient space $S^1/Q$, where $S^1$ is the circle group and $Q$ is the group of
quadrilateral symmetry. It associates to each point of the
surface to be meshed a cross made of four unit vectors orthogonal
with each other in the tangent plane of the surface.
In our case, we compute one cross field per CAD face, with crosses aligned
with the CAD curves bounding the face.

\paragraph{Cross representation}
To represent the crosses, we use the standard unit vector representation
$\vu = (\cos (4\theta), \sin (4\theta)) = (u_1,u_2)$ \cite{hertzmann2000}
where $\theta$ is the angle of one of the four branches with a local
reference frame. Contrary to the naive angle representation $\theta$,
the vector representation $\vu$ is invariant to the quadrilateral symmetries and
suitable for linear finite element interpolation. As the linear combination of unit vectors
is usually not a unit vector, the interpolated values inside triangles are not
a strictly representations of crosses, but the crosses can be recovered easily 
by projecting on the unit circle, \ie normalizing the vector representation.

\paragraph{Finite element discretization}
To discretize the vector field representation of the cross field, we define one
cross at each edge of the triangulation and use the Crouzeix-Raviart
interpolation, as in \cite{beaufort2017}. The unknowns of the cross field problem are the representation
vector field components $u_1,u_2$ at each edge $e_{ij}$. As we want the
cross field to be aligned the surface boundaries, the Dirichlet boundary
conditions are $\vu = (\cos(0),\sin(0))=(1,0)$ at each boundary edge.

\paragraph{Smoothing with the heat equation}

To have a smooth cross field inside the domain, the natural approach is to minimize
the Dirichlet energy: 

$$ \min_{|\vu(\vx)|=1} \int_\Omega \| \nabla \vu \|^2$$

This objective function is non-linear as the crosses live in
the quotient space $S^1/Q$, which is equivalent to the cross
representation being on the unit circle, \ie $\forall \vx \in \Omega, |\vu(\vx)|=1$.
A standard relaxation is to allow the cross representation to leave the unit circle,
but they should stay close to it. Minimizing the Dirichlet energy by directly solving the Laplacian
of the vector representation, \ie $\nabla^2 \vu = {\bf 0}$, is not appropriate
as the values may collapse to ${\bf 0}$ far from the boundary.
One elegant solution to this problem is to
solve a Ginzburg-Landau non-linear equation to penalize the values leaving the
cross manifold \cite{beaufort2017}, but this approach is too expensive for large
problems. A more efficient alternative is the MBO approach \cite{viertel2019},
where one alternates heat diffusion (Eq. \ref{eq:diffusion}) and projection (Eq. \ref{eq:projection}) steps. A further refinement
\cite{palmer2020} developed for 3D cross fields is to start with large diffusion
time-steps and progressively decrease them. In this work, we also use
decreasing diffusion time-steps but we select them according to the mesh size
considerations, and we regroup them in levels to allow for re-use of the matrix
factorization computed by a direct linear solver.  

\begin{align}
    &\text{Diffusion:} &&{\partial \vu \over \partial t} = \alpha \nabla^2 \vu \label{eq:diffusion} \\
    &\text{Projection:} &&\vu \leftarrow \frac{\vu}{\| \vu \|}  \label{eq:projection}
\end{align}

\paragraph{Multilevel strategy}

When dealing with large CAD models, there are usually geometric features at
many scales. Using a uniform mesh that is sufficiently fine to capture the smallest
features is expensive and wasteful, as the cross field is mainly smooth far
from the boundaries. In practice, we use triangulated meshes adapted to the
geometric features, which can lead to edge lengths varying by many order of
magnitudes in the mesh.

Our strategy is to iteratively resolve the cross field at various scales, which
we call levels. Each level is characterized by a different diffusion
coefficient, the $\alpha$ in (Eq. \ref{eq:diffusion}). Intuitively, we
start with large diffusion lengths and progressively decrease to
eventually capture the smallest cross field features. Beside being
suited for non-uniform meshes, it is also a great way to accelerate
computations on uniform triangulations.  This approach is very fast: the cost
of computing an accurate cross field is similar to the cost of solving a few
linear systems, as we are re-using matrix factorizations inside levels.

After generating a triangulated mesh of the whole CAD model, with
mesh sizes adapted to the CAD features, we compute the cross-field 
on each surface independently. As the boundary condition (cross tangency)
is the same on both sides of a curve separating two CAD surfaces, the
solutions will be compatible and we can regroup them at the end.

For a given triangulation $T$ of a surface $\surface$, we pre-compute $N$ diffusion
coefficients that will be used for the $N$ successive heat diffusion levels. At
each level, we successively solve the heat equation and project the solution,
until the cross field no longer changes. In a way, we are solving the cross
field problem (minimizing the Dirichlet energy) for a given resolution.
Typically, the appropriate number of levels is between $5$ and $10$ for
a large model but a lower value around $3$ or $4$ is sufficient for a CAD face.
We chose the initial (largest) and final (smallest)
coefficients as
$$\alpha_\text{initial} = (0.1 * d)^2 \text{ and } \alpha_\text{final} = (3 * h_\text{min})^2$$
with $d$ the length of the surface bounding box diagonal and
$h_\text{min}$ the smallest edge length. The large initial time-step
allows a first coarse diffusion of the boundary conditions and the final small
time-step allows a local smoothing of the crosses at the finer mesh sizes.
Between these extremities, we use linearly interpolated time-steps. While these
choices are quite arbitrary, they are the results of many experimentations and
we have found that they work pretty well on a large variety of CAD models with
features at many scales.

The speed of our approach comes from the re-use of linear factorizations inside each
level. From the heat diffusion equation (Eq. \ref{eq:diffusion}),
we derive the discrete version, using finite element for spatial and backward Euler for time discretization:
$$ M (U_{i+1} - U_{i}) = - \alpha K U_{i} $$
with $M$ the mass and $K$ the stiffness Crouzeix-Raviart matrices. For more
details about the stiffness matrix assembly, one can refer to \cite{beaufort2017} as our
coefficients are the same. With Crouzeix-Raviart, the mass matrix is diagonal
and the coefficients are $M_{jj} = \frac{|t1| + |t2|}{3}$ with $|t1|,|t2|$ the
areas of the triangles adjacent to the edge $j$ \cite{wardetzky2007}.

The heat diffusion system can be re-written as:
$$ (\frac{1}{\alpha} I + M^{-1} K) U_{i+1} = \frac{1}{\alpha} U_{i} $$
At a given level (fixed $\alpha$), the solution $U$ will evolve, so is the
right-hand side, but the matrix part of the system $(\frac{1}{\alpha} I +
M^{-1} K)$ will not change, and so can be factorized one time per level. We can
also note that the sparsity pattern of the system will be the same for all
levels, and thus the preprocessing of the system (reordering, memory
allocation) can be done one time at the beginning.
While the factorization we use is an operation only available with direct
linear system solvers, it can be replaced by an efficient preconditioning when
using iterative solvers.
Putting all things together, we describe the complete process to compute the cross
field in Algorithm \ref{algo:crossfield}. Thanks to the system factorizations,
the total computational cost of our cross field solver is approximately the cost
of a few linear system solves (one per level, typically three or four).

\begin{algorithm}
\caption{Fast heat-based computation of cross field}
\begin{algorithmic}

\State \textbf{Input:} A triangulation of the surface and global parameters:
       $N$ the number of diffusion levels (typically $N=5$) and a norm convergence threshold $c$ (typically $c=1.e^{-3}$)
\State \textbf{Output:} A boundary-aligned smooth cross field $f$ sampled at the edges (Crouzeix-Raviart)

\State \textbf{Procedure:}

\State compute the sparse Crouzeix-Raviart stiffness matrix $K$ 

\State compute the sparse Crouzeix-Raviart mass matrix $M$, which is diagonal

\State process the sparsity pattern of $A = M^{-1} K$ \Comment{Direct solver call}

\State $\alpha_\text{init} = (0.1 * d)^2$, with $d$ the bounding box diagonal length \

\State $\alpha_\text{final} = (3 * h_\text{min})^2$, with $h_\text{min}$ the minimum edge size \

\For{each level $l \in [1,N]$}
   \State $\alpha_l = \alpha_\text{init} + (\alpha_\text{final}-\alpha_\text{init}) * l / (N-1)$

   \State update the linear system matrix diagonal so $A = (\frac{1}{\alpha_l} I + M^{-1} K)$

   \State factorize the matrix $A$ \Comment{Direct solver call}

   \While{$\|U_{i+1} - U_{i}\|_\infty > c$}
       \State update the right-hand side $B=\frac{1}{\alpha_l} U_{i}$

       \State solve the linear system $A U_{i+1} = B$ using the factorization \Comment{Direct solver call}

       \State project the cross coefficients on the unit disk, $U_{i+1} \leftarrow \text{proj}(U_{i+1})$
   \EndWhile
\EndFor

\end{algorithmic}
\label{algo:crossfield}
\end{algorithm}

%
%
%
%
%
%
%
%
%
%
%

Concerning the accuracy and the convergence of the scheme, it should be noted
that we are not interested in reaching the solution of the {\it true}
Ginzburg-Landau continuous formulation.  Obtaining such solution is difficult and expensive, as it
requires a fine mesh and a uniform triangulation, because the Dirichlet energy
tends to infinity at singularities, so it should be uniformly sampled. We are more
interested in cross fields that are useful for our quadrilateral meshing
approach, and for this a smooth cross field with singularities \emph{reasonably
placed} is sufficient.  In this context,
we can apply the same algorithm parameters ($N,c,\alpha_\text{init},\alpha_\text{final}$)
to all models as it is not important if a cross field solution is not
totally converged.

\paragraph{Boundary condition extension}

With uniform meshes, minimizing the Dirichlet energy, while staying close to
the cross manifold with MBO or Ginzburg-Landau, has a tendency to position
singularities close to the boundary. For instance, on a unit circle,
singularities are at the radius 0.85. While this is not an issue when
computing or studying cross fields, it is not practical for quadrilateral
meshing: the associated irregular vertices may be closer to the boundary than
the target edge size, complicating vertex insertion in indirect approaches, or
there is not enough triangles between the singularity and the boundary to have
an accurate parametrization with global parametrization approaches.

To push away the singularities from the feature curves, we use a simple trick:
extend the boundary conditions inside the surfaces. On each triangle touching
a feature curve, we fix the crosses on its three edges. The cross values
are averaged by solving the Laplace equation $\nabla^2 \vu = {\bf 0}$ on this
one-triangle boundary layer.

\subsection{Cross field singularities}
\label{ss:singularities}

To constrain the cavity remeshing (\rsec{sec:cavity}), we use a list of 
singularities extracted from the cross field previously computed with
the heat-based approach.

\paragraph{Detection of singularities.} Cross field singularities are
detected by computing, for each point $\vx$ on the surface, the angle difference along a small 
closed circle $\gamma$ centered on $\vx$. This defines a singularity index \cite{Ray2008}:
\begin{equation}
    \text{index}(\vx) = \frac{2}{\pi} \oint_{\gamma} d\theta (t)
\end{equation}
In contrast to most related work, we use the factor $\frac{2}{\pi}$ instead
of $\frac{1}{2 \pi}$ to match the previously defined (\rsec{sec:topo_quad}) indices of the irregular vertices in the
quadrilateral mesh. When a cross field singularity has
index an of $-1$ with our definition (instead of the usual $-1/4$), its
corresponds to an irregular vertex of index $k=-1$ in the quad mesh, \ie valence five.

In the discrete setting, we extract the oriented edge one-ring $(e_{i1}, ..., e_{in})$
around the vertex $i$ and we compute the sum of the angle differences.
In practice there are three cases: the sum is zero and the vertex is regular or the sum is
one or minus one and the vertex is a singularity of the cross field.

Because of our Crouzeix-Raviart discretization of the cross field (one angle
per edge), cross field singularities may lie on vertices, on edges or on
triangles of the mesh.  When a singularyty lies on a edge, both its
two adjacent vertices have an index different from zero, and when it lies on a triangle, we observe three
adjacent vertices with an index different from zero. For our application it is not necessary
to locate the singularities on mesh vertices, so we explicitly
store the 3D position of the singularities, which can be the middle of a
singular edge or the barycenter of a singular triangle.

\paragraph{Compatibility with quad mesh topology.}
In our experience, cross fields computed with a smoothing scheme and a
continuous vector representation ($\cos(4\theta),\sin(4\theta)$) fail to be compatible with quadrilateral mesh
topology (\reqt{eq:euler_quad_2}) when there are acute corners ($\text{angle} <
45^\circ$) in the CAD face or if the triangulated mesh is not refined enough
along curved CAD curves.

To \emph{correct} the first issue, we artificially add one singularity of index
$1$ to our list of floating singularities at each acute corner of the CAD face.
With this simple addition, there is no need to modify the triangulation or to
recompute a new cross field.  For the second issue, we increase the
sampling of the CAD curves based on the curvature during the initial
triangulation process.

While both techniques usually work, and largely improve our results, there is
no guarantee that the inconsistency between the singularity list and the CAD
face topology will always be fixed. Consequently, for sake of
robustness, \emph{
  we do not assume that the cross field singularity list is
  topologically correct  in the following steps of the meshing
pipeline.} This constitutes a major advantage of our unstructured approach: the ability
to work with cross fields that are \emph{topologically wrong}.

\subsection{Scaling the cross field for integrability}
\label{ss:scaling}

The cross field that has been described in \rsec{ss:heat} is made of unit
vectors. It is not integrable in the sense that if we extract (locally) two
orthogonal vector fields from its branches, they will not commute.  In this
section, we show how to scale the branches in order to ensure local
integrability of the cross field.

\paragraph{Theoretical considerations}
Assume two vector fields $(\vX,\vY)$ 
defined in the tangent plane of $\surface$. We look for a condition on
$(\vX,\vY)$ such that they are \emph{the tangent vectors},
or \emph{directional derivatives}, of a map 
$\vF~:~(u,v) \in \reals^2 \mapsto \surface$. Formally, we want
$\partial_u \vF =  \vX$
and $\partial_v \vF =  \vY$.
The map $\vF$ could be constructed ``naively'' in the following
way. Let us pick a point $\vx = (x,y)\in S$
and choose $\vF (0,0) = \vx$. We first advance ``in time'' of a time increment $u$ along the flow 
$\Psi_{\vX}(\vx,u)$ of  $\vX$. The flow can be seen as follows: if $\vX(\vx)$ is seen as the velocity  
of a fluid at point $\vx$, a massless particle at point $\vx$ will be advected by $\vX$ along
integral lines of $\vX$. After a time increment of $t$, the particle starting at $\vx$ will be
located at $\Psi_{\vX} (\vx,u) \in M$. Note that the corresponding location in the 
parameter plane is $(u,0)$. 
Then we advance in time  of a time increment $v$ along the flow of 
$\vY$ which lead us to point $(u,v)$ in the parameter plane and at point 
$\Psi_{\vY} (\Psi_{\vX} (\vx,u),v)$. A map is only possible if the two flows commute:
\begin{equation}
\Psi_{\vY} (\Psi_{\vX} (\vx,v),u) = \Psi_{\vX} (\Psi_{\vY} (\vx,u),v)
\label{eq:flow}
\end{equation}
From differential geometry, we know that both flows commute if and only if  
the Lie bracket is zero:
$$[\vX,\vY] = J_\vY \vX -  J_\vX \vY = \vz$$
where $ J_\vX$ and $ J_\vY$ are the Jacobian matrices of the vector fields.
In our case, we assume that $\vX$ and $\vY$ have locally the same length: 
$$\|\vX(\vx)\| = \|\vY(\vx)\| = h(\vx) = e^{H(\vx)}$$
Choosing $h=e^H$ is adequate because vector lengths should be strictly
positive so the map ${\bf X}$ is injective.
In 2D, $\vX$ and $\vY$  are two branches of the cross
field that are mutually orthogonal and scaled by $h$. Those vector
fields can thus be parametrized using an angle $\theta(\vx)$:
\begin{equation}
\vX(\vx) = e^{H(\vx)} \begin{pmatrix} \cos(\theta(\vx)) \\ \sin(\theta(\vx)) \end{pmatrix}~~~\text{and}~~~ 
\vY(\vx) = e^{H(\vx)} \begin{pmatrix} -\sin(\theta(\vx)) \\ \cos(\theta(\vx)) \end{pmatrix}. 
\label{eq:st}
\end{equation}
Assuming $\vX$ and $\vY$ of the form \eqref{eq:st}, the Jacobian is
\begin{eqnarray}
J_\vX \vY &=& 
 e^{2H}
\begin{pmatrix}
-H_x \sin{\theta} \cos{\theta} + \theta_x \sin^2 \theta + H_y
\cos^2{\theta} - \theta_y \sin{\theta} \cos{\theta} \nonumber \\ 
-H_x \sin^2{\theta} - \theta_x \cos{\theta} \sin{\theta} + H_y \sin{\theta}\cos{\theta} + \theta_y \cos^2{\theta}  
\end{pmatrix} \nonumber.
\end{eqnarray}
By enforcing $H_x = - \theta_y$ and $H_y = \theta_x$, the jacobian reduces to 
$J_\vX \vY  = e^{2H}\nabla \theta$ and similarly $J_\vY \vX  = e^{2H}\nabla \theta$,
fulfilling the commutativity condition $[\vX,\vY] = J_\vY \vX -  J_\vX \vY = \vz$.
Thus, the condition on the cross field scaling $h = e^H$ for local integrability is
\begin{equation}
\nabla^\perp \theta = \begin{pmatrix}-\theta_y \\ \theta_x \end{pmatrix}= \begin{pmatrix}H_x \\ H_y\end{pmatrix}= \nabla H
\end{equation}

\paragraph{Discretization}

From the cross field computation, we know at each edge $(i,j)$ the angle
$\theta_{ij}$ between one of the cross branches and the edge tangential
direction. We now describe how to compute the scaling factor $h(\vx)$ at each
vertex of the triangulated mesh.

Consider the triangle $(i,j,k)$, we first compute the angle differences
$$
\Delta \theta_{jk} = \text{diff}(\theta_{jk},
\theta_{ij})~~~\text{and}~~~
\Delta \theta_{ki} = \text{diff}(\theta_{ki},
\theta_{ij}).
$$
Operator $\text{diff}(\theta_{jk},\theta_{ij})$ is not equivalent to
a simple difference of angles because (i) angle $\theta_{jk}$ is
measured on a different system of coordinates as $\theta_{ij}$ and 
because (ii) the cross branches have a $4$-symmetry which
implies that $\theta_{jk} \equiv \theta_{jk}+k_2\frac{\pi}{2}$ is the 
same cross for any $k_2 \in Z$. We choose $k_1$ and $k_2 \in Z$ in such a way that 
$$
\theta'_{ij} = \theta_{ij} + k_1{\pi \over 2} \in
\left[0,\frac{\pi}{2}\right]~~~,~~~
\theta'_{jk} = \theta_{jk} - \alpha_j + k_2{\pi \over 2} \in
\left[0,\frac{\pi}{2}\right]
$$ 
with $\alpha_j$ the angle between the tangential vectors of the edges $(i,j)$ and $(j,k)$, 
and then define
$$\text{diff}(\theta_{jk},\theta_{ij}) = \left\{
\begin{array}{lcr}
\theta'_{jk} - \theta'_{ij}& \text{if} & |\theta'_{jk} - \theta'_{ij}|
                                         \leq \frac{\pi}{4}\\
\theta'_{jk} + \frac{\pi}{2} - \theta'_{ij}& \text{if} & \theta'_{jk} - \theta'_{ij}
                                         < -\frac{\pi}{4}\\
\theta'_{jk} - \frac{\pi}{2} - \theta'_{ij}& \text{if} & \theta'_{jk} - \theta'_{ij}
                                         >  \frac{\pi}{4}
\end{array}
\right.
$$

It is thus possible to compute $\nabla \theta$ as a constant vector
of the tangent plane in every triangle. Computing $H$ can be done
by solving 
\begin{equation}
\min_{H} \int_{S}| \nabla H - \nabla^\perp
  \theta|^2d\vx
\label{eq:fem}
\end{equation}
using $P^1$ finite elements.
The scaling factor is simply obtained as $h(\vx) = e^{H(\vx)}$.
An example of conformal scaling is illustrated on \rfig{fig:gparam_vs_qqs}.a.

Equation \eqref{eq:fem}
only involves $\nabla H$ so $H$ is defined up to a constant $C$. It is
indeed possible choose $C$ in such a way that the final mesh
contains about $N$ quads. We actually know that 
$$N \simeq \int_S {1 \over h^2 (\vx)}d \vx = \int_S e^{-2(H (\vx)+C)}
d \vx = e^{-2C} \int_S e^{-2H (\vx)} d \vx.$$
Thus,
$$e^{-2C} = {N \over \int_S e^{-2H (\vx)} d \vx}~~~\rightarrow~~~C =
-{1 \over 2} \ln \left( {N \over \int_S e^{-2H (\vx)} d \vx}\right).$$

\paragraph{Comparison with related work} 
The scaling that we are a computing is a conformal scaling
associated to the cross field. It has been computed in the past directly from the
cross field singularities by solving a linear system \cite{bunin2007,ben2008}.
The difference is that we do not \emph{robustly} know the singularities.
By solving the least square problem (\reqt{eq:fem}) which only involves
cross gradients, we always have a scaling 
which is usable even if our cross field is not globally consistent (\eg wrong singularities).

\section{Size map and curve quantization}
\label{sec:sizemap_and_curve}

To generate a mesh, it is necessary to have some kind of size map specifying
the target size of the elements on the domain. When an explicit scalar size
field is not used, it can be defined implicitly by simply using a uniform target
size, or prescribed sizes on some CAD entities (corners, curves) with a maximal
uniform size far from them.  If there are small CAD curves in the model, or
small prescribed sizes, the size transition to the coarser regions is usually
constrained by a maximal mesh gradient which is a parameter (often hidden) of
most meshers.

Most complex CAD models have features of many scales and
using the smallest feature size as a uniform target size is not
practical as it would lead to meshes with an excessive amount of
elements. In our experience, letting the meshing
algorithms, which usually have not been designed with mesh size transitions in
mind, to just \emph{do their things} and hope for the best is not optimal
as it often leads to poor quality quad meshes.  To robustly produce high-quality meshes
on generic models, we have to use a smooth size map to control the transition
from small features to much larger ones.

The conformal scaling associated to the cross field (\rsec{ss:scaling})
induces a size map (up to a constant) which takes into account the
cross gradients (\reqt{eq:fem}). This size map is critical to
be able to generate square-shaped quads in the smooth regions of the
model. Yet, it ignores small CAD features.
Thus, trying to build a quad mesh with such discrepancy between the
forced mesh edges (from the CAD) and the (implicit or explicit) size map used
by the quad surface mesher leads to poor quality elements, even if the process is
robust. The ideal solution would be to have a cross field which respects the
CAD features and mesh size prescriptions, such that its conformal
scaling would be naturally compatible with the CAD geometry. Unfortunately,
this problem is still open.

Here, a smooth size map is computed from the CAD
feature sizes and is blended with the size map from the conformal scaling
by taking the minimum of both size maps. The final size map
allows for smooth transitions from the smallest CAD features to the coarse
regions where it is dominated by the conformal scaling.

\subsection{Size map from small CAD features}
\label{ss:small}

Small CAD feature is a generic term used for the regions where
a small mesh size is required to capture the geometry with well-shaped
elements. A small CAD curve is a small feature, as well as two
CAD objects that are close to each other.

In this work, a \emph{minimal size} field $s^{\min}(\vx)$ is built
for taking into these small CAD features.
$s^{\min}$ is a nodal field whose support is the triangular mesh
that has been used for computing the cross field. Its values are
initially computed on CAD curves. Mesh size $s^{\min}(\vx)$ at a vertex $\vx$
that belongs to a CAD curve $\curve_j$ (i) is never larger than the
actual length of $\curve_j$ and (ii) is never larger that the distance
between $\vx$ and the closest  CAD curve $\curve_k$ that is
not topologically adjacent to $\curve_j$. The field values in $s^{\min}$ are
subsequently propagated to the internal nodes of the CAD surfaces,
ensuring a smooth gradation. 


Technically, the propagation consist in a Dijkstra-like algorithm:
the size $s^{\min}$ is updated on every edge $(\vx_1,\vx_2)$ of the
triangulation (including edges classified on CAD curves)
in order to limit the size gradient to a prescribed
value $g_{\max}$:
\begin{equation}
    s^{\min}(\vx_2) = \min(s^{\min}(\vx_2), s^{\min}(\vx_1) + g_{\max}\; d(\vx_1,\vx_2))
    \label{eq:one_way}
\end{equation}
This simple approach works actually well in practice. For more
information on mesh size vs. CAD features, the reader can refer to
\cite{bawin2020}.

\begin{figure}
\begin{center}
    \includegraphics[width=\textwidth]{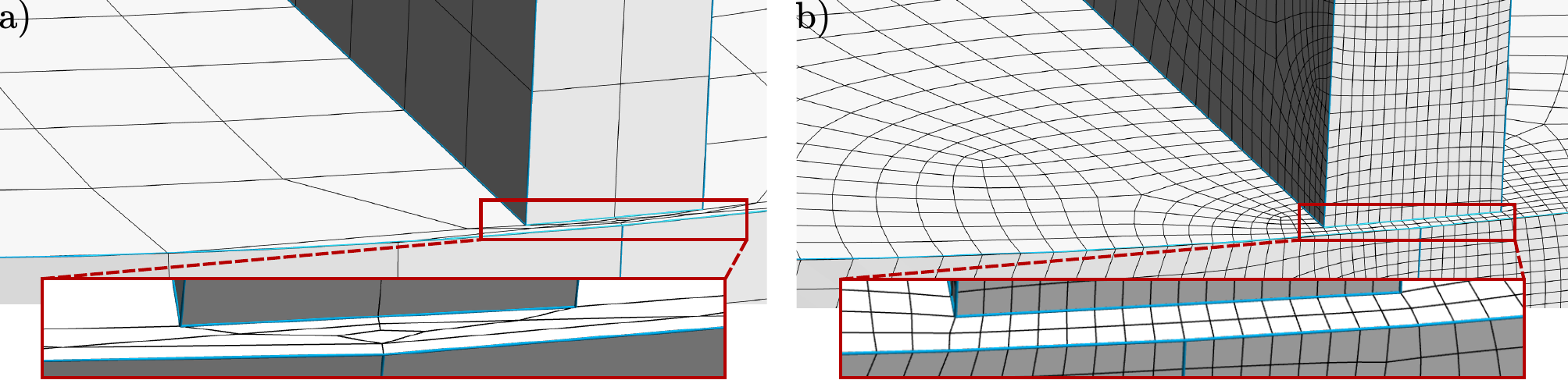}
\caption{
    Focus on a small CAD feature of the \emph{Block} model. 
    Without adaptation of the size map (a), there are low quality elements
    around the small feature. With size map adaptation (b), there are
    small high-quality elements around the feature and irregular vertices
to make the transition to coarser regions.}
    \label{fig:sizetr}
\end{center}
\end{figure}

\subsection{Global size map}
\label{ss:gsizemap}

To build a global continuous and smooth size map $s(\vx)$, we combine the minimal
size field $s^{min}(\vx)$ with the size maps $s_i^{cf}(\vx)$
associated to the cross field conformal scaling (\rsec{ss:scaling}) computed
on each CAD face $\surface_i$.

The conformal scaling fields (\rsec{ss:scaling}) are defined up to a constant
$C_i$ that can be tuned in order to have $N_i$ quads on the surface
$\surface_i$. Starting from a global target number of quads $N$, we distribute
the quads on each CAD face based on their areas, \ie $N_i =
\frac{\text{area}_i}{\text{total area}} N$. Another advantage of this simple
approach is that we can easily take user preferences into account. If a
user want a specific number of quads on a CAD face, we can use this value
instead of the one proportional to the area.

The conformal scaling size maps are not continuous at
the CAD interfaces. Consider a point $\vx$ on a CAD curve common to the faces
$\surface_i$ and $\surface_j$, we have $s_i^{cf}(\vx) \neq s_j^{cf}(\vx)$
because the cross field gradients, and thus the conformal scaling, are not
equal on both sides of the curve.  To make a continuous and smooth union of the
$s_i^{cf}$ fields in a global one $s^{cf}$, we apply a similar smoothing process
as before (\rsec{ss:small}). Initialize the union of the size map, taking the
minimum on the shared vertices (CAD corners and curves): $s^{cf}(\vx) =
\min_i(s_i^{cf}(\vx))$. Then apply a one-way smoothing (\reqt{eq:one_way}) with
the maximal gradient $g_{max}$.

To blend the small feature size map and the conformal scaling, we simply take
the minimum everywhere:
$$s(\vx) = \min(s^{cf}(\vx),s^{min}(\vx)) $$

A local example of our size map taking the small CAD features into account is
shown on \rfig{fig:sizetr}, where small quads are used in the region where two
feature curves are close to each other. This figure is a zoom on a small part
of the larger \emph{Block} model (\rfig{fig:block}), which have lot of small features and
where significant size map transitions are necessary in many regions.

\subsection{Curve quantization}
\label{ss:curve}

We call quantization the process of going from continuous fields (cross field,
size map) to a quadrilateral mesh which is characterized by discrete
quantities, such as the number of points on a model curve or the number of
quads in a model face.  As we are following the robust Gmsh's ``surface to
volume'' philosophy, the curve quantization is important in our pipeline.

The process of first meshing the model curves then the model surfaces,
constrained to the curve discretization, works well for the generation of
triangulations (or tetrahedrizations) but it is not optimal for
quadrilateral meshing (or hexahedral meshing) as the quad topological
constraints (chords) are global and go through the model curves. For
instance, imagine a simple rectangular face for which the ideal edge
numbers, according to the continuous cross field induced size map, are $[9.4,
4.1, 10.6, 3.9]$ on the four sides. A simple quantization by integer rounding
would choose respectively $[9, 4, 11,4]$ edges. While a quadrilateral mesh with
this quantization exists, it is not structured as it must include one pair of
+1/--1 irregular vertices to make the transition from $9$ edges on one side to
$11$ edges on the opposite side.

Finding a \emph{good} curve quantization for quad meshing, for a generic CAD
model with CAD faces of arbitrary topology, is a complicated and totally open
problem, which has been mostly ignored in the literature.  If we restrict ourselves
to quadrilateral CAD faces (with T-junctions), there is the global quantization in
\cite{campen2015}, but in our case we do not have such quadrilateral layout.

For the current paper, our quantization of CAD curves remains quite simple.  We
use the integer-rounded value of the ideal number of edges computed by
integrating the global size map (\rsec{ss:gsizemap}), except for curves which
are on quadrilateral CAD faces, where we impose equality on the opposite sides.
This non-optimal quantization leads to many necessary dipoles (pair of +1/--1 irregular
vertices) in the final quad meshes of the different CAD faces. Eliminating them
requires to change the CAD curve 1D meshes, which is out of the scope of the current paper.

\paragraph{Size map integration on curves}

Consider a CAD curve $\curve$ parametrized
by $t$ varying from $a$ to $b$. The \emph{floating-point} number of edges on
the curve according the size map is:
$\bar{n}_e = \int_{a}^b dl / s(\vx)$ with $dl=\norm{x'(t)}dt$. To get a integer number
of edges $n_e$, we use a simple rounding: $n_e = \text{round}(\bar{n}_e)$.
The parameter $t_i, i \in [1,n_e-1] $ associated to the $i$-th interior point of the curve
is such that:
\begin{equation}
\int_{a}^{t_i} \frac{dx}{s(\vx)} = \frac{i}{n_e} \int_{a}^b \frac{dl}{s(\vx)} 
    \label{eq:1D_integration}
\end{equation}
This integral equation can be solved via numerical integration, by adding
values along the curves until the sum is equal to $i \bar{n}_e / n_e$, with linear
interpolation between the samples. 

To mesh the curve $\curve$, we compute the vertex positions $\vx_i, i \in [1,n_e-1]$
by computing the parameters $t_i$ with (\reqt{eq:1D_integration}) and evaluating
the CAD curve parametrization: $\vx_i = f(t_i)$. With this approach, the points
are well placed on the curves according to the size map, making smooth transitions
from the small CAD feature regions to the coarser ones, where smaller but 
useful variations come from the cross field conformal scaling.

\paragraph{Topological constraints on quadrilateral patches}

Integrating the size map does not take into account the particularities of
quad mesh topology, such that the quads are organized in topological
chords (dual loops of adjacent elements). In this work, we adjust the
number of edges $n_e$ on the opposite sides of quadrilateral CAD faces.

Consider a CAD face whose boundary is made of
four CAD curves $(\curve_1,\curve_2,\curve_3,\curve_4)$, we force the number of
points to be equal on the opposite edges, \ie $n_{e1}=n_{e3}$ and
$n_{e2}=n_{e4}$, except if the integrated values are very different.
When two adjacent quadrilateral CAD faces share a curve $\curve_c$,
the value $n_{ec}$ must be the same on both faces. This means
the equality constraints propagate in the CAD model.
To resolve the propagation,
we build the topological chords associated to the quadrilateral CAD faces. A
chord is made of topologically parallel CAD curves, which all receive the same
fixed number of points, that we compute by averaging the ideal values
previously computed on each curve.

With this simple propagation, the chords propagate only if two adjacent
quadrilateral faces share a single curve which is one of their four sides. We
do not propagate across T-junctions.  Our quantization applies only on the
connected subsets of conformal quadrilateral patches of the CAD model.  Dealing
with T-junctions is more complicated and corresponds to the quantization phase
of \cite{campen2015}, for which there is not always a solution
(\rsec{ss:motivation}).


\section{Unstructured frontal quadrilateral mesh generation}

\label{sec:unstructured}

Now that the 1D meshes of the CAD curves have been generated, it's
time to move to the initial quad-meshing of the CAD surfaces.
The all-quad mesh is generated by following a robust three step
approach: (1) generate a triangulation with right-angle triangles
using the methodology in \cite{delquad}, (2) combine some 
triangles into quads to form a quad-dominant mesh and 
(3) apply midpoint subdivision to produce an all-quad mesh. 
We also describe the geometry smoothing and untangling techniques
(\rsec{ss:geometry}) which are used extensively in our pipeline.

\subsection{Frontal point insertion}
\label{ss:frontal}

A scaled cross field ${\bf f} (\vx)$ has been generated where the size
of the crosses is the global size map  $s(\vx)$ (\rsec{ss:gsizemap}).
Mesh vertices are created on every
face $\surface$ in a frontal fashion, starting from its boundary
$\partial \surface$. Vertices of $\partial \surface$ are
added in a priority queue. Vertex $\vx$ at the top of the queue
tries to add four new vertices in the domain in the 4 directions of
the local cross field $f(\vx)$ and at a distance $s(\vx)$. New vertices are only
added (i) if they lie inside $\surface$ and (ii) if they are not to close
to any existing vertex. When a vertex is successfully added, it is inserted
at the end of the queue and the procedure continues until no new vertices
can be generated. The points are iteratively inserted in the triangulation
with a Delaunay kernel, as described in \cite{delquad}. This mesher
is close to classic frontral Delaunay triangular meshing \cite{rebay1993},
except that the goal is to build right-angled triangles in agreement with
the cross field.

\subsection{Quad combination and midpoint subdivision}
\label{ss:combination}

The triangulation is transformed into a quad-dominant by combining
pairs of triangles into quads. As triangles are allowed, there is no
need for advanced matching technique such as Blossom-Quad \cite{blossom}
and we simply use a greedy selection. All the merge candidates (two triangles $\to$ one quad)
are computed and sorted by the geometric quality (\rsec{ss:geometry}) of the resulting quads, 
weighted by the alignment with the cross-field. Then the quads are
iteratively selected, as long as their quality is superior to a minimal
threshold (SICN $> 0.1$ in our case).

The all-quad mesh is obtained by subdividing all quads into four sub-quads and all
triangles into three sub-quads. This subdivision is called midpoint subdivision and 
is topologically equivalent to the Catmull–Clark subdivision surface technique.
New vertices are added at the centers of each edge and each element.

To accurately represent the CAD model geometry, it is best to project the new midpoint vertices
on the CAD curves and faces. However, this projection is not robust in the sense that it may
create invalid quads (negative quality). This difficulty is similar to the one faced when
doing CAD snapping for high-order meshing, and is not easy to solve.
In our pipeline, we iteratively try to snap all midpoint vertices while monitoring the quality
of the adjacent elements, and we revert the projections which lead to negative elements.
Even if rejected projections are rare and definitively not ideal, this check is necessary
to ensure the geometric validity of the final mesh when dealing with complex CAD models.
For better geometric fidelity, one could decrease the size map locally to better capture
the geometry and restart the meshing process of the CAD curves and surfaces in the neighborhood.

An example of quad mesh produced by this procedure is shown in
\rfig{fig:gparam_vs_qqs}.2.b. It is actually a fairly good unstructured all-quad mesh, yet
containing an excessive number of irregular vertices.

\subsection{Geometry optimization}
\label{ss:geometry}

Geometrical smoothing of a mesh is the process of finding \emph{good} vertex
positions given a fixed mesh topology. Ideally, we would like the positions
to maximize some quality functional, but such formulation are usually
too expensive to be solved globally and we have to resort to more efficient techniques. 
Smoothing of unstructured quadrilateral
meshes is still a challenging problem in our opinion. Simple and fast
techniques, such as Laplacian smoothing and its variants, often produce tangled elements
(negative quality) while more advanced non-linear optimization techniques are
computationally too expensive to be used on large sets of elements.  For CAD
surfaces, an additional expensive operation is the projection of a 3D point on
the surface, which is typically performed many times during optimization. As an
approximation, it is possible to project on a triangulated representation of
the geometry, but even this approximation is expensive when applied millions of
times.

In this work, we apply a great deal of topological remeshing (\rsec{sec:topological_quad})
and each time we use smoothing techniques to determine the geometry before
validating or rejecting the remeshing. Concretely, this means that we are applying
smoothing all the time, on many more elements that there are in the initial or final
mesh. For instance, if there are $100k$ vertices in the quadrilateral mesh, we will
probably smooth millions of vertex positions. While smoothing is always expensive
in any mesher, the computational cost is particularly critical in our approach and
we have to employ fast techniques and fall-back to expensive ones only locally and
in last resort. Before and after any smoothing operation, we compute the element
qualities and we revert the vertex positions if the quality decreased.

\paragraph{Quality metric}
For consistency with other \emph{Gmsh} algorithms, we use the Signed Inverse
Condition Number (SICN) metric to evaluate the geometric quality of the quads.
It is the inverse of the condition number in Frobenius norm of the mapping from the
quad to the regular square.  This metric, which is also called \emph{shape quality}
\cite{verdict}, has a behavior similar to the scaled Jacobian: the range is
$[-1,1]$ with negative values for invalid elements and a maximum value of $1$
for squares (and rectangles for the scaled Jacobian).  As with the Jacobian, 
the SICN value is not constant inside a
quad, so we use the minimum. In practice, most mesher sample the quad quality
at the four corners and take the minimum. Assuming the 
quad corners are $\vx_i, i \in [1,4]$ and $N_i$ are the CAD surface unit normals, the scaled Jacobian and SICN values at the i-th corner 
are respectively \cite{verdict}:
\begin{equation}
    SJ_i = \frac{\vL_{i-1} \times \vL_{i}}{\norm{\vL_{i-1}}\norm{\vL_{i}}} \cdot \vN_i,
    \quad
    SICN_i = 2 \frac{\vL_{i-1} \times \vL_{i}}{\norm{\vL_{i-1}}^2+\norm{\vL_{i}}^2} \cdot \vN_i,
    \; \text{ with } \vL_i = \vx_{i+1}-\vx_{i}
\end{equation}
When one edge length $\norm{\vL_i}$ tends to zero (geometrical edge collapse), 
the scaled Jacobian will not tend to zero but to the sine of the angle at the corner
(as $\vL_{i-1} \times \vL_{i}=\norm{\vL_{i-1}}\norm{\vL_{i}} \sin(\theta)\ \vn$).
This observation have very practical implications: 
when sampling the quality at the four corners, a quad mesh may have many almost-collapsed
edges (\eg $1e^{-12}$) but still a high minimum scaled jacobian (\eg $0.7$).
On the other hand, the SICN quality metric, also sampled at the four corners, will
tend to zero with edge collapse. In our experience, the discretized SICN 
should be preferred to the scaled Jacobian when evaluating the quality of isotropic quadrilateral meshes.

\paragraph{Laplacian smoothing in CAD parameter domain} When we are dealing with
a parametrized CAD model, we have a valid $uv$ trimmed parametrization of each
CAD face.  To smooth the vertices inside a cavity $C$, which can be a whole CAD
face, we extract continuous $\vuv \in \reals^2$ values for each boundary $\vx \in \partial C$.
One must be careful with the parametrization jumps (\eg periodic seam on a
cylinder). Sometimes it is not possible to have a continuous boundary in the
parameter domain (\eg around the pole of a sphere parametrization), and we move
to the next smoothing options. Once we have a continuous $\vuv$ on the boundary,
we simply solve the unweighted Laplacian equations $\Delta u = 0$ and 
$\Delta v = 0$, 
where each internal value is the arithmetic average of its edge-connected
neighbors. It is important to use the arithmetic average and a direct linear
solver to respect the maximum principle and avoid inverted elements in the
parameter domain. Depending on the distortions in the CAD parametrization,
this technique may both produce high-quality or poor elements.
To retrieve the 3D positions, we then apply the CAD face mapping
$\vx = f(\vuv)$. Depending on the resulting element qualities, we keep or we reject
this smoothing. 
The huge advantage of this approach is that there is no surface
projection and simply two linear solves, making it a very fast smoother.
Even if the quality may be poor due to distortion in the CAD mapping, it is 
still a good initial guess since all elements are usually untangled.

\paragraph{FDM Winslow kernel on regular vertices} 
Winslow smoothing \cite{winslow1966} is the industry-standard technique used
for structured grid mesh generation \cite{thompson1998}. Given a fixed
boundary, the idea is to solve the Winslow non-linear elliptic PDE $\Delta_x \vu =
\mathbf{0}$, where $\vu(\vx)$ are the coordinates in a certain computational
space and $\vx$ are the coordinates in the physical space.
The advantage of this approach is that the two coordinate components are
coupled and the resulting quads are nicely shaped, with some orthogonality
enforced even under large distortions. By applying a finite difference (FDM)
discretization to the Winslow equation \cite{knupp1999}, we can derive a local smoothing kernel for
regular vertices in the quadrilateral mesh. Assuming that $(\vx_1, ..., \vx_8)$
are the ordered vertices of the stencil around the regular vertex $\vx$, its
new position is given by:
\begin{equation}
    \begin{split}
    \alpha_0 (\vx_1+\vx_5 - 2 \vx)
    + \alpha_1 (\vx_3 & +\vx_7  2 \vx)
    - \frac{1}{2} \beta (\vx_2 + \vx_6 - \vx_4 - \vx_8) = 0 \\
    \text{with } 
        \alpha_0 = (\vx_3-\vx_7) \cdot& (\vx_3-\vx_7), \quad
    \alpha_1 = (\vx_1-\vx_5) \cdot (\vx_1-\vx_5), \quad
    \beta = (\vx_1-\vx_5) \cdot (\vx_3-\vx_7)
    \end{split}
\end{equation}

This smoothing is moving the vertex outside of the CAD surface. 
We project back on it by finding the closest point
on a triangulated representation of the CAD surface (with a kd-tree for faster
spatial queries). If we want a more accurate projection, such as in the last
steps of the smoothing, we ask the projection to the CAD geometric kernel,
which in our case is OpenCASCADE. 

\paragraph{Angle-based kernel on irregular vertices} On irregular vertices,
we use the angle-based smoothing kernel of \cite{akram2021}, where the idea
is to move the vertices on the bisector of the one-ring. As the above Winslow
kernel, it works well in regions where the mesh is not too constrained but
it may create inverted elements in complex configurations. An alternative
for irregular vertices is the unstructured Winslow FDM kernel \cite{knupp1999},
but it is more complicated and would suffer from the same issues.

\paragraph{Smoothing loop}
The smoothing kernels, followed by projection on the CAD,
are applied iteratively to all vertices of
the current cavity, until the displacement is smaller than a criteria or the
maximal number of iterations is reached. This iterative process is slower than
the uv-Laplacian smoothing, because of the non-linearity of the kernel
and of the surface projections, so we generally use it without waiting for
convergence. It should be notated that these kernels may produce tangled
elements, especially close to concave CAD features or in very irregular
regions. For this reason, we compare the element qualities before and after,
and roll back if necessary.

\paragraph{Non-linear optimization} Some local configurations
are difficult to untangle and smooth, or even impossible. This is generally due to a mix of constraining
CAD features and mesh irregularity, such as for the disk quadrangulation remeshing
(\rsec{sec:disk_quad}). In these tricky situations, we use the non-linear 
untangler of Mesquite \cite{mesquite} on very small patches of quads. This
optimization can fail or may convergence too slowly, so in practice we use
a time limit of $1$s. Failure to untangle is not a big issue in our
pipeline because we start from a geometrically valid unstructured mesh by
construction, so untangling is only encountered during remeshing and we
can simply rollback if the result is tangled.

\section{Topological quadrilateral meshing}
\label{sec:topological_quad}

Unstructured quadrilateral meshes generated frontally (\rsec{sec:unstructured})
typically have a large excess of irregular vertices (\eg \rfig{fig:M3}.c.). 
To transform the mesh into a quasi-structured quad mesh (\eg \rfig{fig:M3}.d.), 
we need to remove most of the irregular vertices while preserving the ones that important
to accommodate the model topology and geometry. We achieve this topological
improvement by extracting \emph{cavities} of quads in the unstructured mesh and by
replacing them by \emph{more regular} quad meshes.
We act at three different
scales: the cavity may be an entire CAD face when it is topologically simple
(\rsec{sec:pattern}), the cavity may be a convex patch of quads in a CAD face
(\rsec{sec:cavity}), or the cavity may contain only the quads adjacent to
a vertex (\rsec{sec:disk_quad}). For large cavities
(\rsec{sec:pattern} and \rsec{sec:cavity}), we look
for replacement meshes in a list of high-quality predefined patterns (\rfig{fig:patterns}).
For local cavities, we look for replacements in the list
of all possible disk quadrangulations (\rsec{sec:disk_quad}).

Each time we apply a topological operation (meshing or local remeshing), it is
very important to verify that the geometric quality of the mesh is
satisfactory. We always apply the geometric smoothing techniques defined in
\rsec{ss:geometry}. If the geometry is invalid,
or if the quality decreased too much compared to the initial configuration
after remeshing, we cancel the mesh replacement.

\paragraph{Common definitions}
A \emph{cavity} $C$ is a set of quads forming a
simply-connected part of the mesh. The boundary of a cavity $\partial
C$ is a closed polyline. Consider a vertex $\vx \in \partial C$. Let us
call $n_{\vx}$ the number of quads adjacent to $\vx$, which are
also the quadrilateral valences. We distinguish
$n^{in}_{\vx}$ the number of quads adjacent to $\vx$ that belong to
$C$ and $n^{out}_{\vx}$ the number of quads adjacent to $\vx$ that do
not belong to $C$. We have $n_{\vx}=n^{in}_{\vx} + n^{out}_{\vx}$.
It should be noted that we usually only consider the quads in the current CAD face $\surface$, 
not the global quadrilateral mesh of the model.

The vertex $\vx \in \partial C$ is a convex corner of the cavity if and only if $n^{in}_{\vx}=1$.
It is a concave corner of the cavity if the vertex is strictly inside the CAD face, 
$\vx \notin \partial \surface$, and if $n^{out}=1$.

By splitting the cavity boundary $\partial C$ at the convex and concave
corners, we form $m$ regular \emph{sides}, which respectively contain 
$N_1, ..., N_m$ edges.


\subsection{Pattern-based quadrilateral meshing}
\label{sec:pattern}

Given a cavity with $m$ sides (containing $N_1,..,N_m$ edges), it is often
(not always) possible to topologically find a quadrilateral mesh to fill
the interior from a list of predefined patterns (\rsec{ss:matching}). 
Such quadrilateral meshes (\rfig{fig:patterns} bottom) are
the anisotropic subdivision of coarse quadrilateral meshes
(\rfig{fig:patterns} top). By anisotropic subdivision, we mean that each 
topological chord (dashed lines in \rfig{fig:patterns}) can be subdivided
independently of the others. We use this constrained topological 
approach to mesh whole CAD faces when their topology is simple (\rsec{sss:simple_face}),
skipping the frontal unstructured mesher, or to iteratively improve patches of quads
in the initially unstructured mesh of a CAD face.

\subsubsection{Matching patterns}
\label{ss:matching}

\begin{figure}
    \begin{center}
        \includegraphics[width=\textwidth]{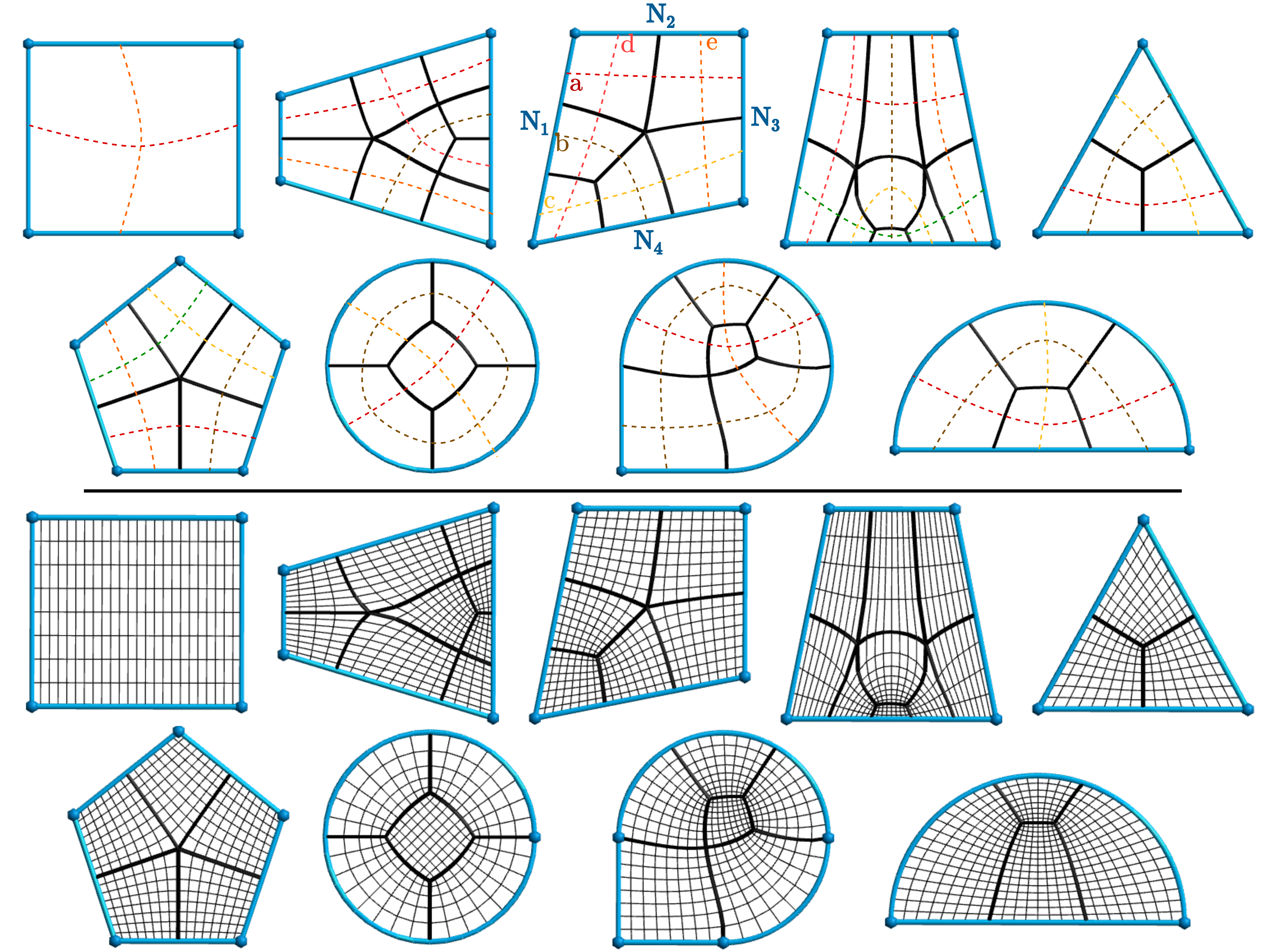}
        \caption{Pattern-based topological quad meshing. The patterns are defined
            by coarse quadrilateral meshes (top). The topological chords (dashed lines)
            can be subdivided independently to build refined quadrilateral meshes (bottom).
        }
        \label{fig:patterns}
    \end{center}
\end{figure}

Consider the cavity with side edges $(N_1, ..., N_m)$, we check if there is a matching
with a quad pattern with $m$ sides in our pre-computed pattern list (\eg \rfig{fig:patterns}).
For each pattern $\mathcal{P}$, we denote $(s_1, ..., s_c)$ the number of edge subdivision of
the $c$ topological chords of the patterns (dashed lines \rfig{fig:patterns}).
There is a quadrilateral mesh if it possible to find strictly positive chord subdivisions 
$s_j > 0, j \in [1,c]$
such that on each side, the sum of the subdivisions of the chords included in the side is equal to
the side number of edges: 
\begin{equation}
\forall i \in [1,m], \quad \sum_{j \in [1,c]} w_{ij} \; s_j = N_i
    \label{eq:ILP}
\end{equation}
where $w_{ij}$ is the number of times the chord $j$ in inside the side $i$. In practice,
$w_{ij}$ is equal to $0$ (the chord is not on the side), $1$ (the chord go through the pattern) or $2$
(the chord starts and finish on the side, turning inside). This linear system
may be undetermined if there are more chords (columns) than sides (rows). We
are not only interested in a solution to the linear system, we want one that is integer and strictly
positive because it corresponds to the chord subdivisions of the pattern. We also
want the values to be well balanced so the output quadrilateral mesh is not
highly distorted. This behavior can be obtained by finding the solution that is
the closest to an ideal subdivision of the chords, defined by equi-balancing the
values on each side. This corresponds to a non-linear objective function 
where we minimize a distance, \eg $E = \sum_j {(s_j - x_j)^2}$ with $x_j$ the
ideal subdivisions.

To solve such system, with linear constraints and a quadratic objective function,
the ideal approach is to use a suited Mixed-Integer solver. From a practical
point of view, such solvers are large software dependencies and would be overkill
for our problem. As our system is usually very small, we have adopted a much
simpler solution: compute the integer row echelon form of the system matrix, 
start from an ideal guess (by balancing the values) and use a depth-first search
to find a solution satisfying all the constraints by rolling up in the echelon form matrix.
However, even if the system is small, the
exhaustive search that happens when there is no solution is too slow for practical use.
So we abandon the search after a few hundred attempts. While this approach may sound
 naive, we observe that it works well in practice as the solution, if there is one, is close
to the ideal subdivisions (initial guess).  It should be noted that when looking for match with
a pattern, one must check all the rotations of the sides $[N_1, ..., N_n]$,
clockwise and counter-clockwise.

The advantage of this integer formulation is that it is generic and works on any quadrilateral pattern.
The user can add a new pattern by simply specifying the coarse quadrilateral mesh. The
construction of the sides, chords and linear systems are all handled automatically.
While it is possible to construct thousands of patterns by enumerating the quadrangulations
of the disk (\rsec{sec:disk_quad}), it should be noted that most of them do not lead to high-quality
quadrilateral meshes. For this reason, we restrict ourselves to the patterns
of \rfig{fig:patterns} that we selected manually.

For a practical example, consider a rectangular CAD face with respectively
$[N_1,N_2,N_3,N_4]$ edges on its sides. It can potentially match with the first
four patterns of \rfig{fig:patterns}. For the first pattern, the condition
is obviously equality of the subdivisions on the opposite sides ($N_1=N_3$ and $N_2=N_4$).
Let consider a more interesting case: the third pattern. The sum of
the chord subdivisions on each side gives the system:
\begin{align}
    a+b+c &= N_1 \nonumber \\
    d+e &= N_2 \nonumber \\
    a+c &= N_3\nonumber \\
    b+d+e &= N_4 \nonumber
\end{align}
We can see that $b$ must respect $b = N_3 - N_1$ and $b = N_4-N_2$, which
is a quite natural condition if we look at the dual chords (\rfig{fig:patterns}). 
On the other hand, there is no solution for this pattern if $N_3 - N_1 \neq N_4 - N_2$.
Assuming the above condition is respected, the balance between the respective subdivisions
$a,c$ and $d,e$ are left to the minimization of the objective function, or to the
depth-first search with our technique. 

\paragraph{Related work} 
The above integer formulation is not new and has been
introduced for patterns with a single interior vertex in \cite{bunin2008}.
A general ILP formulation for more complicated patterns has been developed in \cite{takayama2014}
and is equivalent to ours (\reqt{eq:ILP}).

\paragraph{Geometry} 
The integer formulation is purely topological and does
not take geometry into account. A very interesting extension would be to
find a way to incorporate geometric information from the cavity boundary into the 
objective function, in order to find optimal chord subdivisions.
In our pipeline, the mesh geometry is obtained by smoothing (\rsec{ss:geometry})
with a fixed boundary, and the new quadrilateral mesh is only kept if its quality
is better than the initial one.

\subsubsection{Application to simple CAD face meshing}
\label{sss:simple_face}

In a complicated CAD model (\eg \rfig{fig:block}), there are usually lot of CAD
faces but many of them have the topology of simple polygons (\eg triangles,
rectangles, pentagons). Given such polygon and the number of edges on each of
its sides, it is often possible to directly find a quadrilateral mesh with our
topological matching technique (\rsec{ss:matching}). 

In practice, we loop over all CAD faces and check if they have the same
topology ($\chi$ and convex corners) as some of our patterns. When the topology
is identical, we search for a subdivision of the pattern which matches the
number of edges on each side of the CAD face, previously determined by the
curve quantization step (\rsec{ss:curve}). This matching exists if there is a
strictly positive solution to the integer linear problem (\reqt{eq:ILP}).
Once the quadrilateral mesh topology is determined, we can find the vertex
positions by employing our quad mesh geometry smoother (\rsec{ss:geometry}).
We only keep the quad mesh only if the geometric quality is satisfactory, and
if it is not we fall back to the unstructured approach.

\begin{figure}
    \begin{center}
        \includegraphics[width=\textwidth]{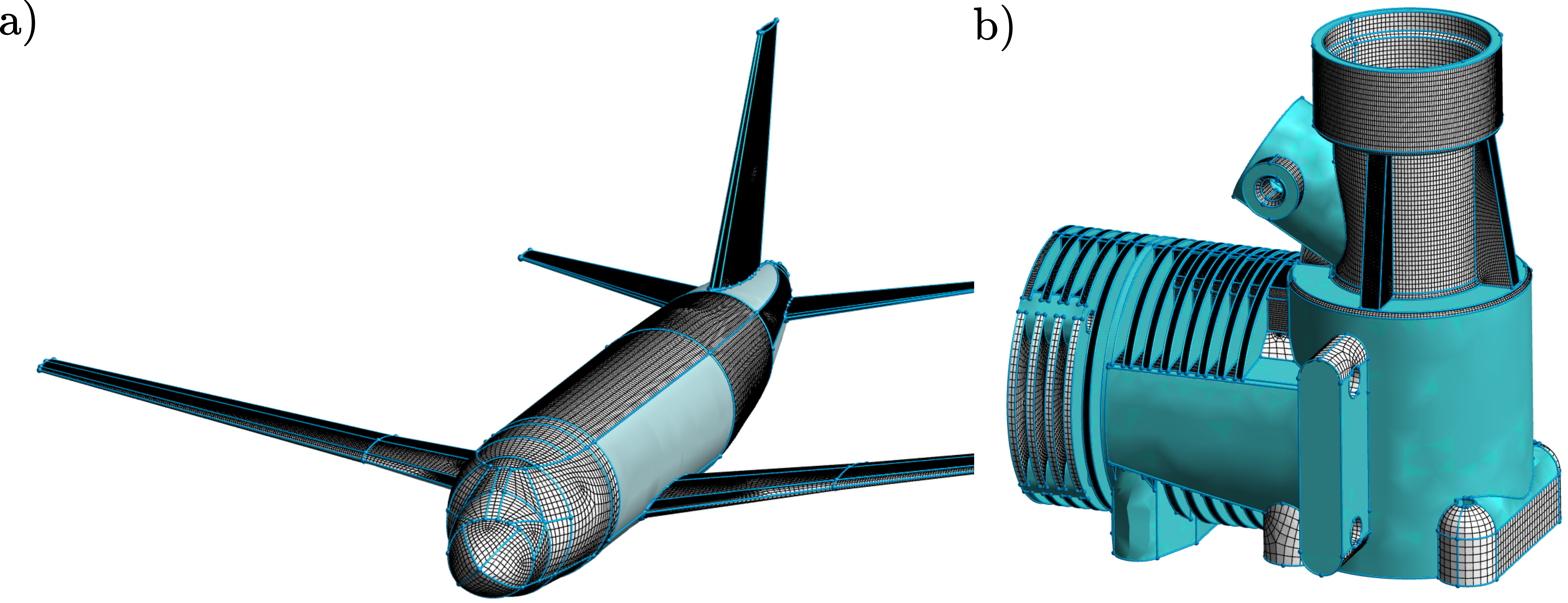}
        \caption{Pattern-based quadrilateral meshes of $75/83$ CAD faces on the A319 model (a) and
        $419/533$ CAD faces on the Block model (b).}
        \label{fig:patterns_results}
    \end{center}
\end{figure}

\paragraph{Examples} To see the usefulness of this pattern-based pre-meshing in real-life
situations, we can look at the results on two CAD models. The
\emph{A319 model} is made of 83 CAD faces, 275 CAD curves and 287 CAD corners
and the \emph{Block model} is made of 533 CAD faces, 1584 CAD curves and 1044
CAD corners. After the curve meshing (\rsec{ss:curve}), we apply our
pattern-matching on all CAD faces and we are able to directly build the
quadrilateral meshes of respectively $75/83$ and $419/533$ CAD faces, as shown
in \rfig{fig:patterns_results}. These results are very interesting because it
means that we are able to build high quality meshes
on most of the simple CAD faces in a short amount time, for which there is
no need to use more computationally expensive techniques.

\subsection {Quasi-structured topology with cavity remeshing}
\label{sec:cavity}


On the remaining CAD faces, we start from the unstructured
quadrilateral mesh (\rsec{sec:unstructured}) and we improve its topology by locally
remeshing cavities with the pattern-matching technique (\rsec{sec:pattern}).
Our objective is to iteratively reduce the number
of irregular vertices in order to reach a quasi-structured topology. Consider a
CAD face, there are four sources of irregular vertices: (i) the face
topology (see \reqt{eq:euler_quad_2}), (ii) the face geometry, (iii) the non-ideal
quantization of the curves (\rsec{ss:curve}) and (iv) the non-ideal behavior of the
frontal mesher (frontal collisions, size map not adapted to CAD
features or cross field gradients, etc). The irregular vertices from (i) and (ii) are captured as
cross field singularities (including the ones we artificially added at acute corners).
In the current paper, we work on removing the irregular vertices
(iv) caused by the unstructured quadrilateral mesher, while preserving the
irregular vertices (i) and (ii) which are necessary to have a \emph{good}
quadrilateral mesh in the end.
It should be noted that the irregular vertices (iii) caused by the non-ideal quantization cannot be
removed by considering only one CAD face, but the whole model
quadrilateral mesh should be taken into account. We leave this last improvement for future
work as it is definitively not trivial: the global quadrilateral mesh is potentially
non-manifold (\eg multiple volumes) and the CAD features must be strictly preserved
while applying global operations across multiple CAD faces.

To remove the unnecessary irregular vertices, we build convex topological cavities
in the quadrilateral mesh, which contains three or more irregular vertices, and
we replace them by \emph{more regular} quad meshes 
(the patterns defined in \rsec{sec:pattern}).
This process is illustrated
on \rfig{fig:growth}, where a pentagonal cavity with five valence $3$ ($k=1$)
and six valence $5$ ($k=-1$) irregular vertices ($\sum k = -1$) is replaced by
a pentagonal patch with only one valence 5 irregular vertex ($k=-1$). The key
to our approach is to not apply this process blindly but in a specific way to
preserve the irregular vertices matching cross field singularities.

The idea of remeshing cavities with more regular patterns is not new
\cite{bunin2008,takayama2014} and has been exploited in quad meshing 
\cite{jaal,marcias2015,verma2015}. The
difficulty is in choosing which cavities to remesh, because 
uncontrolled greedy remeshing will remove irregular
vertices that were important for mesh quality, leading to highly distorted meshes. For instance,
a rectangular cavity may eliminate a valence three singularity close
to a convex CAD curve and a valence five singularity close to a concave
CAD curve, which were both useful irregular vertices used to accommodate
the model geometry.

We call \emph{singularities} the irregular vertices that we 
preserve, which mainly come from the cross field singularities,
and we call \emph{unnecessary} irregular vertices the ones
we want to eliminate. The singularities are used to control
and constrain the growth of the remeshable cavities that absorb 
the unnecessary irregular vertices.

\subsubsection{Growing remeshable cavities}

\begin{figure}
    \begin{center}
        \begin{tabular}{cccc}
            \includegraphics[width=0.22\textwidth]{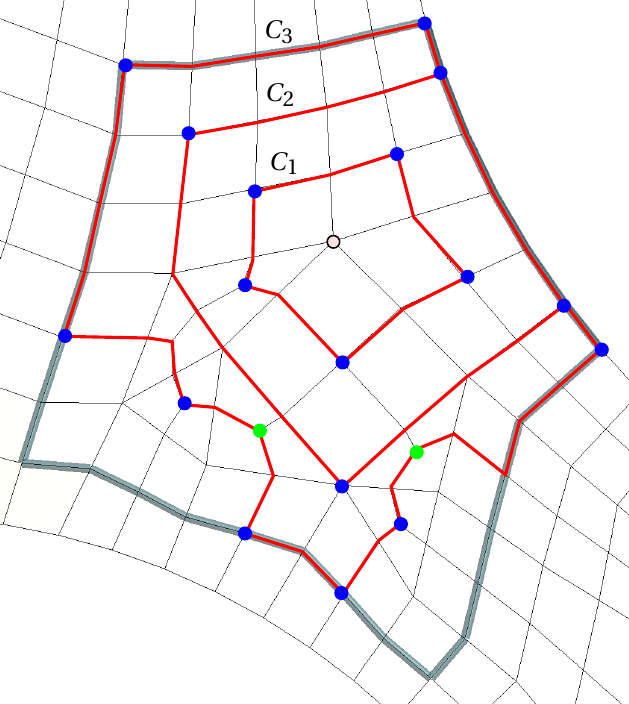}&
            \includegraphics[width=0.22\textwidth]{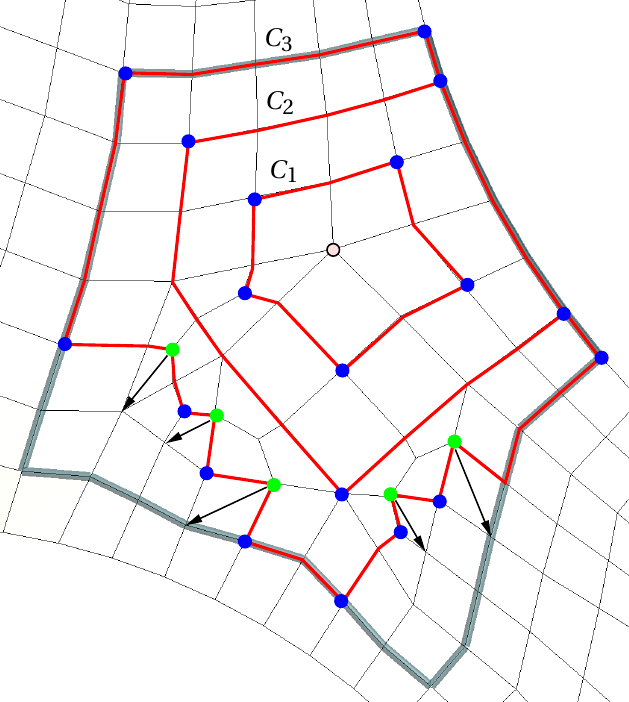}&
            \includegraphics[width=0.22\textwidth]{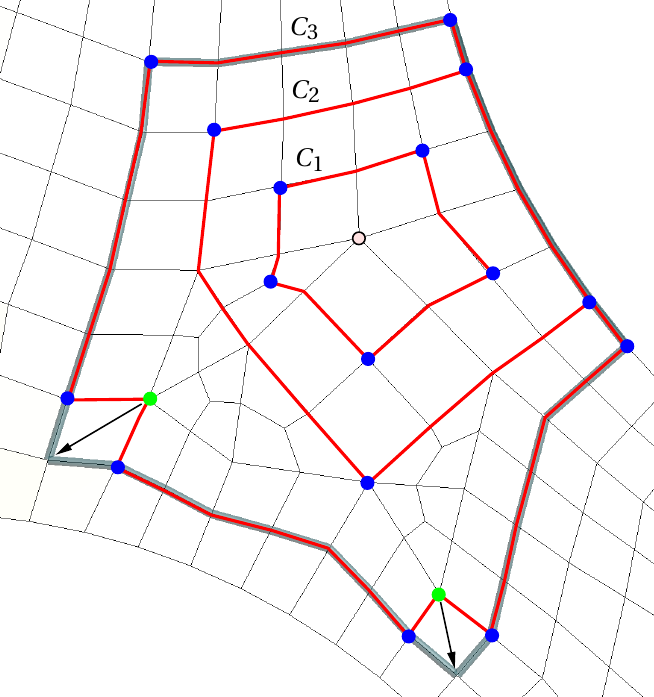}&
            \includegraphics[width=0.22\textwidth]{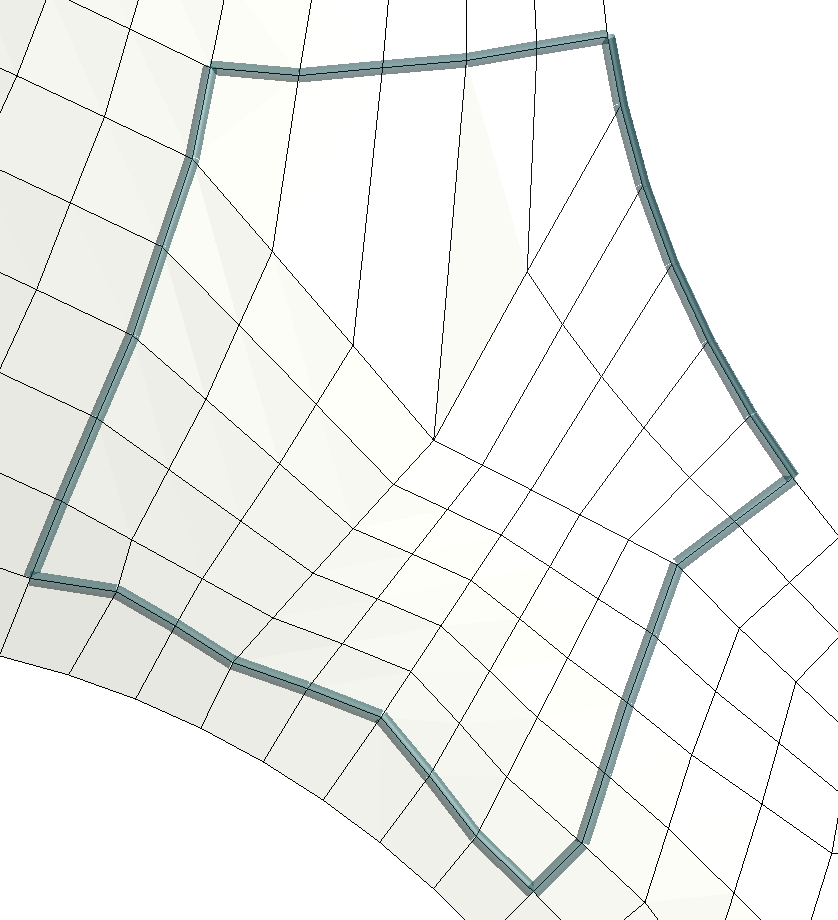}\\
            (a) & (b) & (c) & (d)
        \end{tabular}
    \end{center}
    \caption{Growing a cavity around one vertex of index $-1$ (in
    pink). Convex corners are in blue and concave corners are in
    green. The remeshed cavity (d) has one irregular vertex instead of eleven.
\label{fig:growth}}
\end{figure}

To iteratively improve the mesh of a CAD face $\surface$, we grow convex
cavities (patch of quads) and replace them. To grow a convex cavity $C$, we
start from an initial simply connected set of quads $C_0$ (\eg quads adjacent to
a vertex) and we iteratively add quads adjacent to the cavity. To ensure the
convexity of the cavity, if one of the vertex $\vx$ on the boundary $\partial
C$ is concave, \eg $\vx \notin \partial\surface$ and $n_{\vx}^\text{out}=1$, we
add the adjacent exterior quad in priority. 
During the growth, we also
ensure that (i) a singularity (flagged irregular vertex to preserve) is never
added to the interior of the cavity and (ii) a CAD concave corner is never
fully surrounded by the cavity ($\vx$ concave CAD corner, $n_{\vx}^\text{out}=0$). 
This process is illustrated in \rfig{fig:growth}, where a pentagonal cavity
is growing starting from a valence-5 singularity.
Each time the growing cavity is convex and has absorbed new irregular vertices,
we check if it is remeshable by solving the matching problem (\rsec{ss:matching}) with
the target replacement patterns. If it is remeshable, we store it as the \emph{last remeshable}
cavity. We either stop if we want a minimal cavity, or we continue the growth if we want 
a maximal cavity.

\subsubsection{Cavity remeshing strategy}
\label{ss:strategy}

The choice of cavity seeds and the order in which cavities are grown and remeshed
matters. When a triangular or a pentagonal cavity built around a singularity is 
remeshed, removing pairs of $+1/-1$ irregular vertices, it produces
quads with only one irregular vertex (associated to the initial singularity),
but the new singularity topological position is \emph{translated} compared
to the initial one. On \rfig{fig:growth}.d., we can observe that the new
irregular vertex is moved closer to the top side (distance of one edge instead
of three initially). Another way to interpret this is that when a pair of $+1/-1$ irregular vertices
is eliminated via a singularity, it will attract it or push it away, depending
of the pair orientation and the singularity index.
This behavior has important consequences: if a mesh singularity is used a lot to
absorb all other irregular vertices, it may end up far away from the initial
cross field singularity, leading to important distortions in the mesh geometry.
So in order to produce a high-quality quasi-structured mesh, we must adopt
a cavity remeshing strategy that will minimize the movements of the
singular vertices and keep them close to their initial position in the cross field.

On a given CAD face $\surface$, we apply the following steps as long as
there is improvement (\ie in a while loop):
\begin{enumerate}[\hspace{1cm}Step 1:]
    \item Starting from unnecessary irregular vertices, grow and remesh maximal
        rectangular cavities that do not contain singularities and that match
        the regular grid pattern (first one on \rfig{fig:patterns}).  The goal
        is to eliminate the opposed irregular $+1/-1$ pairs without distorting
        the mesh.
    \item Starting from the singularities (index $k=+1$ or $k=-1$), grow and
        remesh minimal triangular or pentagonal cavities that contain the
        starting singularity and the non-zero minimum number of unnecessary
        irregular vertices while keeping the target index (\ie $\sum k = -1
        \text{ or } +1$). The goal is to eliminate the irregular vertices
        while distributing the mesh distortion on all the singularities.
    \item Re-apply Step 1, in case some new regular cavities have been \emph{unlocked} by
        the previous step.
    \item Starting from unnecessary irregular vertices, grow and remesh
        minimal rectangular cavities that do not contain singularities
        and that match one of the four rectangular patterns of \rfig{fig:patterns}.
        The goal is to enable size transitions in the mesh while reducing
        the number of irregular vertex pairs. To avoid large geometric distortions,
        we use minimal remeshable cavities and we verify that the mesh quality
        is not degraded too much after smoothing.
\end{enumerate}

This process is the core of our quasi-structured topological improvement.
\rfig{fig:gparam_vs_qqs}.2. illustrates the absorption of irregular
vertices by the cross singularities. For a more complicated example,
we can look at the \emph{Block} model illustrated in \rfig{fig:block},
where the zooms show the kind of mesh size transitions which are produced by
the rectangular pattern cavity remeshing.

\subsection {Local remeshing with disk quadrangulations}
\label{sec:disk_quad}

Unstructured quad meshes constructed in an indirect fashion may contain a 
few very irregular vertices, that we call defects. The goal of this
local remeshing is to eliminate them before applying the larger cavity
remeshing (\rsec{sec:cavity}). In the interior of a
CAD face, we say that a vertex is very irregular if its quadrilateral valence
is more than 6, \ie its index is $k \leq -2$. On the CAD curves, we say that a
vertex is a defect if its valence is different from two on the adjacent CAD
faces. On CAD corners, we want to enforce the ideal valence which is deduced
from the associated angle in the CAD faces, \eg valence three for a concave
$270^\circ$ corner. We can also enforce user-prescribed corner valences
with this technique, which can be useful when hand-crafting boundary layers
for specific applications.
The goal of the local remeshing step is to eliminate the defects (valence
different from the allowed range) by changing the topology locally.

\begin{figure}
\begin{center}
    \includegraphics[width=4.cm]{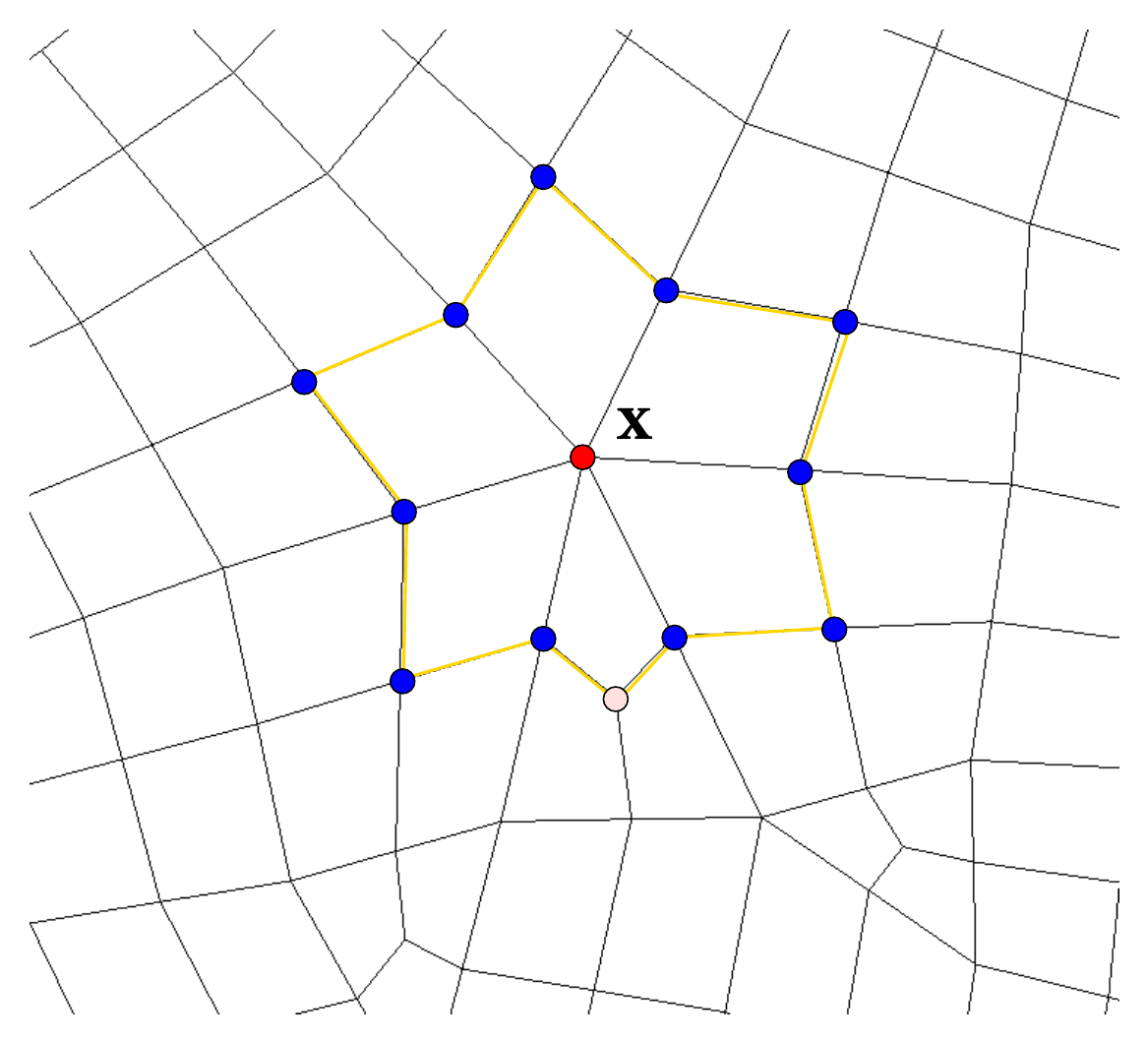}
    \includegraphics[width=4.cm]{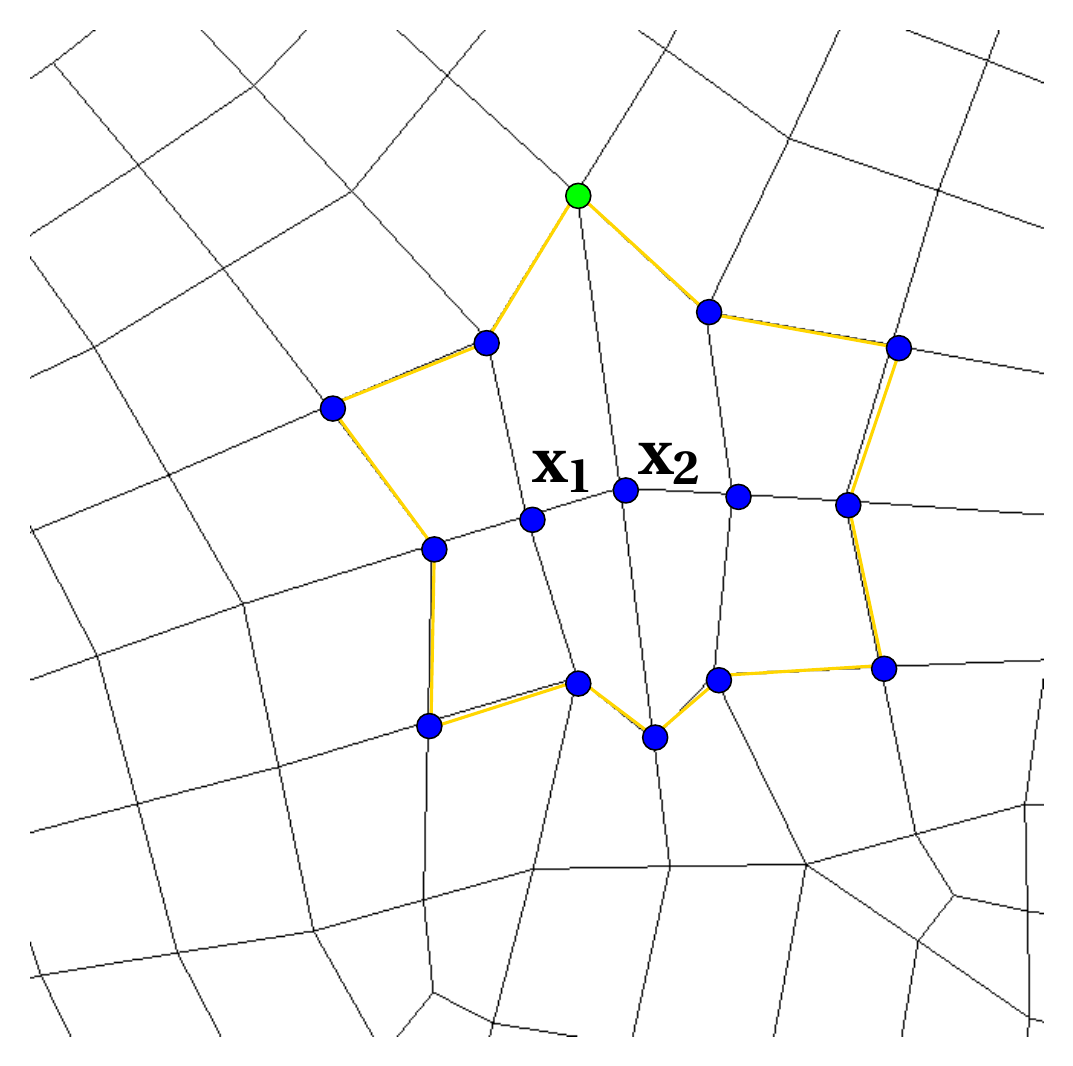}
        \caption{Removing a very irregular vertex $\vx$ (in red) of index $k=-2$.
            Blue dots are regular vertices of index $k=0$. The greed dot has an
        index $-1$ while rose dot has index $+1$. \label{fig:6}}
    \end{center}
\end{figure}

When forming a cavity with the quads adjacent to a vertex, we are building
a topological disk. We can replace the interior of the cavity with any
quadrilateral mesh as long as we do not change the cavity boundary, which
is a closed polyline. Our idea is simple: browse all the possible disk quadrangulations
(\rsec{ss:disk_quadrangulations}) and use the best one, according to a \emph{quality} criterion (\rsec{ss:disk_remeshing}).
An example
of elimination of a valence six vertices is shown on \rfig{fig:6}.

\subsubsection {Quadrangulations of the disk}
\label{ss:disk_quadrangulations}

\begin{figure}
\begin{center}
    \includegraphics[width=10.3cm]{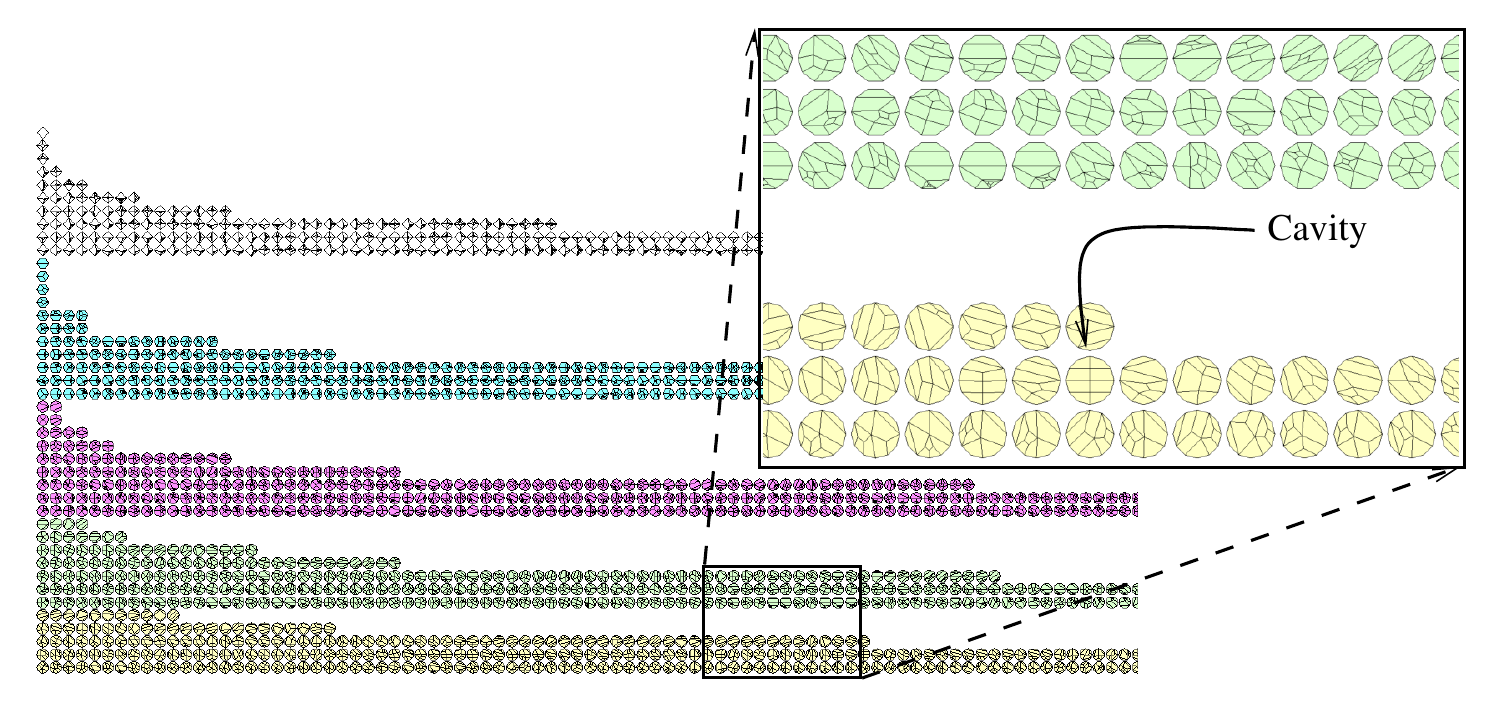}
\caption{All quadrangulations of the disk bounded by $4$, $6$, $8$,
  $10$ and $12$ vertices with $0$, $1$, $2$, $3$ and $4$ internal vertices. \label{fig:patches}}
\end{center}
\end{figure}

The exhaustive list of disk quadrangulations can be built by recursively
applying the three edge flips that add a quad to a current quadrangulation,
starting initially from a single quad.  This is a simple 2D version of the
algorithm used to build shellable hexahedral meshes of the sphere 
in \cite{verhetsel2019}. Our implementation is open-source
\footnote{Software to enumerate disk quadrangulations: https://git.immc.ucl.ac.be/reberol/disk\_quadrangulation}.
The only technical difficulty is in detecting equivalent quadrilateral meshes
that have different vertex ordering. For this we use the open-source
library \emph{Nauty} \cite{nauty} to compute the canonical labeling of the
edge graphs. In our meshing software, we only store the list of the disk quadrangulations
up to a certain size in a large table.

For practical meshing, the only disk quadrangulations that are interesting are the
ones with valence three, four, five inside (indices $1,0,-1$) and valences one, two, three
on the boundary (indices $1,0,-1$). On \rfig{fig:patches}, we show the disk quadrangulations
with up to $12$ vertices on the boundary and up to $4$ vertices in the interior.
These configurations are more general that the patterns previously used (\rsec{sec:cavity})
because they include concave vertices on the cavity boundary.

\subsubsection {Local remeshing}
\label{ss:disk_remeshing}

Consider a cavity $C$ that we want to remesh, with its boundary $\partial C$ defined
by the polyline $[\vx_1, ..., \vx_b]$. We want to select the \emph{best}
replacement in the list of disk quadrangulations with $b$ vertices 
on the boundary, including all rotations clockwise and counter-clockwise
of the patterns.  As the cavity is immersed in a larger quadrilateral
mesh, the quadrilateral valence outside the cavity associated to
each vertex of $\partial C$ is fixed: $[n_{\vx_1}^\text{out}, ..., n_{\vx_b}^\text{out}]$. The future
quad valences inside the cavity $[n_{\vx_1}^\text{in}, ..., n_{\vx_b}^\text{in}]$ are parameters
of each disk quadrangulation. We also know the ideal valences $[\bar{n}_{\vx_1}, ..., \bar{n}_{\vx_b}]$
in the global quadrilateral mesh according to their position in the CAD face,
\eg $\bar{n}_{\vx_j}=4$ inside if regular, $\bar{n}_{\vx_j}=2$ on a curve, $\bar{n}_{\vx_j}=3$ on a concave corner.

The best disk quadrangulation replacement is the one which minimizes the \emph{irregularity}
of the cavity after inclusion in the complete quadrilateral mesh:
\begin{equation}
I_r = \sum_{j=1..b} (\bar{n}_{\vx_j} - n_{\vx_j}^\text{out} - n_{\vx_j}^\text{in} )^2
      + \sum_{k \in \{\text{inside}\}} (4-n_{\vx_k}^\text{in})^2
\end{equation}

If there are multiple ideal quadrangulations ($I_r=0$), we choose the one with the minimal
number of vertices. But in practice, it is not always possible to find ideal replacements,
and it is important to not introduce new defects in the mesh. So we restrict the allowed
integer ranges of the $n_{\vx_j}^\text{in} $. For a CAD face interior vertex
($\bar{n}_{\vx_j} =4$), the allowed range is such that 
$3 \leq n_{\vx_j}^\text{in} + n_{\vx_j}^\text{out} \leq 5$. 
For a CAD curve vertex,
we typically force the regularity: $n_{\vx_j}^\text{in} + n_{\vx_j}^\text{out} = 2$. 
For a CAD corner, we also force the ideal valence: $n_{\vx_j}^\text{in} + n_{\vx_j}^\text{out} = \bar{n}_{\vx_j}$.
Because the problem is very constrained when the CAD topology
and geometry is locally complex,
it is not always possible to satisfy
all constraints (even if we are considering thousands of replacement
candidates, thus the difficulty of quad meshing). To still improve the
quadrilateral mesh, we introduce priorities: it is better to improve the
corners, then the curves, then the interior. We allow the creation of curves
and interiors defects in order to fix the corners, and then we allow the
creation of interior defects when fixing the curves. Such relaxation is
necessary to not get stuck early in the improvement process.

\paragraph{Geometry} 
As the disk quadrangulation are purely topological, the
actual mesh geometry is computed with our smoothing and untangling technique
(\rsec{ss:geometry}). If we fix the boundary of the cavity, there is usually no
geometrically valid solution. This is due to the fact that the quad topology
typically changes significantly while the boundary stays a small polygon: there
is not enough room to untangle and optimize the elements.  To relax the
geometric problem, we extend the smoothing cavity to the quads adjacent to the
topological cavity, and we apply the optimization on this enlarged cavity, with
its boundary fixed. If there is still no satisfactory geometric solution
(\eg tangled quad), we cancel the current disk quadrangulation remeshing.

\paragraph{Limitations}
In practice, most of the defects are eliminated, but sometimes it is not
possible to remove some of them when the region is too constrained (\eg almost
all vertices of a cavity on CAD curves). With luck, they will be eliminated
later by larger convex cavity remeshing (\rsec{sec:cavity}) but there is no
guarantee. That being said, what we call defect are local configurations which
are totally acceptable in the broader context of unstructured quad meshing techniques. 

\section{Results and discussions}
\label{sec:results}

\begin{figure}
    \begin{center}
        \includegraphics[width=1\textwidth]{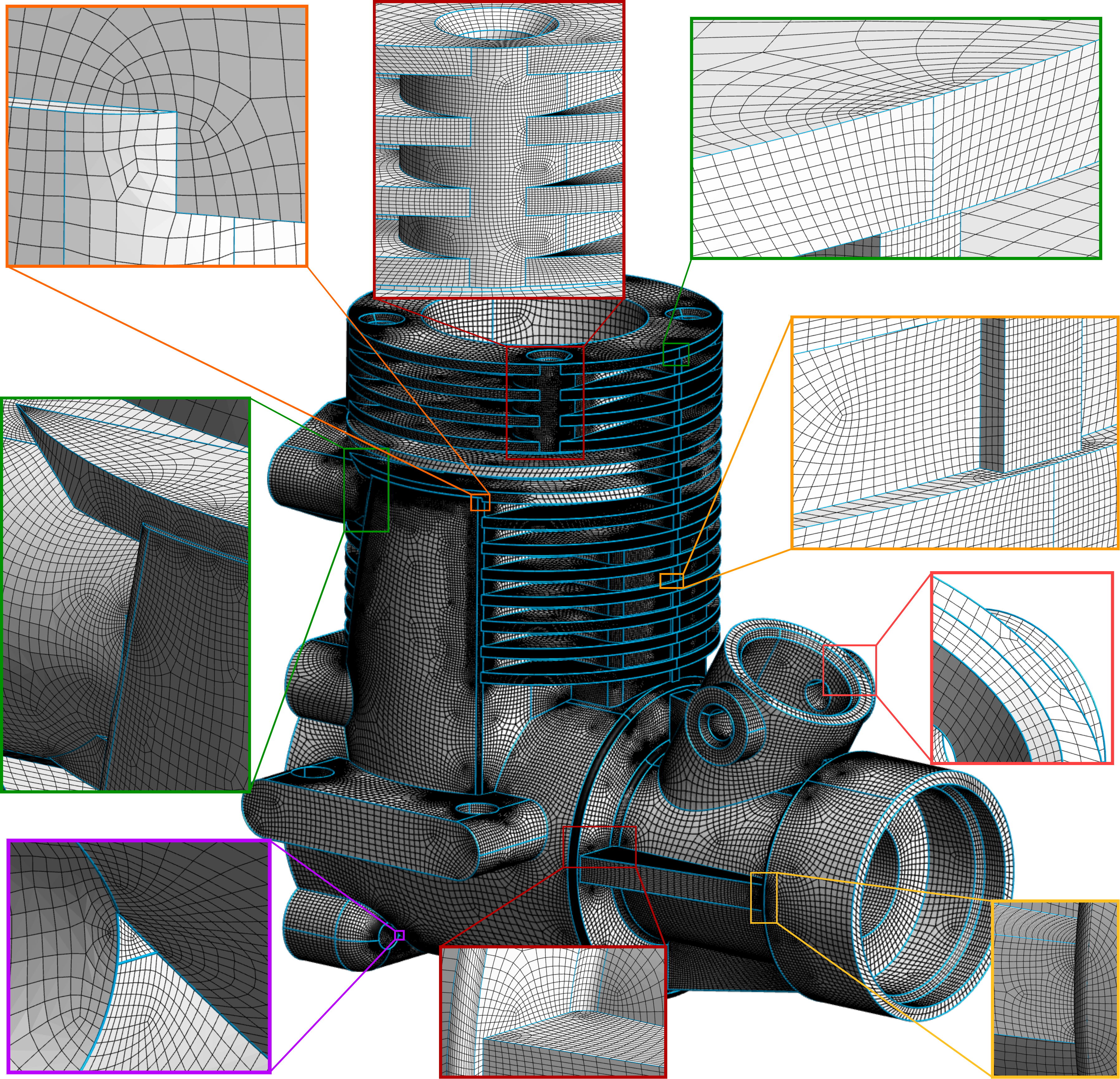}
        \caption{Quasi-structured quad mesh (261.5k vertices, 261.6k quads) of the \emph{Block} model (533 faces and 1584 curves).
            The average SICN quality is 0.87 and the minimum is 0.11. The initial unstructured
            quad mesh was generated in 58 seconds and the quasi-structured improvement
            took 33 seconds, both with 4 CPU cores on a laptop. The number of irregular vertices was reduced
        from 14.4k to 3.6k.}
        \label{fig:block}
    \end{center}
\end{figure}

To validate our approach and confirm its robustness, we applied it on
many CAD models from existing databases (\rsec{ss:database}).
In the next section (\rsec{ss:discussion}), we discuss the strengths
and weaknesses of the approach and its positioning relative to existing
techniques in the literature.

\subsection{CAD databases}
\label{ss:database}

\paragraph{Test cases} We applied the end-to-end meshing pipeline to the $114$
models of the MAMBO dataset \cite{mambo} and to $1000$ models the ABC dataset
\cite{abc}. As the ABC dataset is huge (one million models), we randomly
selected one thousand not-too-simple models for which it was possible to generate triangulations and
extract a single CAD volume. This pre-selection is necessary because many
models are very simple (\eg a few CAD primitives), or are not valid, or contain lot of independent components.
For all models, we applied a global sizing parameter of $0.1$ on mesh
edges, \ie edges ten times smaller than the \emph{initial} very coarse CAD sampling
in the triangular Delaunay mesher.
In our pipeline, a size map is then produced to adapt to CAD feature locally,
but the mesh resolution in the final mesh is largely controlled by this single
parameter. It should be noted that using the same resolution parameter for 
all CAD models is very far from ideal, as the resulting meshes are either
too coarse or too fine relatively to the CAD geometry, compared to what
a user would choose. We still followed this procedure because it
allows large scale automation and it is a good way to observe the
mesher behavior in non-optimal conditions.

\paragraph{Results}
We were able to automatically build geometrically valid quad meshes on
the $114/114$ models of the MAMBO dataset and on $988/1000$ models of the ABC
dataset.  In the $12$ failure cases in the ABC dataset, two are topologically
valid quad meshes but with tangled elements, the minimum SICN quality being
respectively $-0.008$ and $-0.11$. The invalid quads are due to inverted
triangles in the quad-dominant mesh, which is too coarse in some regions. The
ten other failures are due to either (a) issues in the Delaunay point insertion
used for the initial triangulation and frontal quad-dominant mesher or 
(b) performance issues (exceed the 8GB RAM or timeout after 5 hours) in the quasi-structured topological improvement when
the size map is too fine.  The cases (b) typically correspond to elongated models with
large thin regions. Our automatic size map leads to very small elements and many
millions vertices, many of them irregular, in the CAD faces. Our topological
improvement strategy is not adapted in these situations because too many
cavities (literally millions) are built, smoothed and remeshed, which is
computationally too expensive. The above failure cases can be fixed by
using a more appropriate global sizing parameter or by manually building
a size map better adapted to the thin regions.

\begin{table}
\begin{center}
    \begin{tabular}{ | l | c | c | c | c | c | c | c | c |}
    \hline
    Dataset (\#)  & \#C & \#F  & Q min & Q avg & \#vert & \% irreg. init.         & \% irreg. & time init. / improv. \\ \hline
    MAMBO (114) &    46   &  18    &    0.53    &  0.98  & 20.2k & 1.24\% & 0.27 \% & 9.6 sec / 4.6 sec \\ \hline
    ABC (990) &    98   &  38    &    0.44    &  0.96  & 49k & 2.2\% & 0.74 \%   & 15.3 sec / 140.8 sec * \\ \hline
    \end{tabular}
\end{center}
    \caption{Average values over all dataset models. 
        \#C and \#F are the number of CAD curves and faces.
        Mesh statistics are
        computed on the final quasi-structured mesh, except
        for the initial percentage of irregular vertices, which is
        computed on the initial unstructured quad mesh. The
    quality metric Q is the signed inverse condition number (SICN).
Time init. is the elapsed time to build the unstructured quad mesh starting
from the CAD and time improv. is the elapsed time to make it quasi-structured,
both computed with a single core. (*) The average run-time of 140.8 sec is high due to slow
outliers (seven models took more than one hour), the median value is 5.9 seconds.}
    \label{table:results}
\end{table}

\paragraph{Statistics} Topological and geometrical statistics, along with
figures, are available for each model in the supplementary material. Average
values for some quantities are listed in Table \ref{table:results}.
Statistically, the quads are of very high quality and the topological
improvement reduces the number of irregular vertices by a factor
of three (ABC) or five (MAMBO). An important proportion of the remaining irregular vertices are due to 
large gradients in the size map. This
is caused by the global sizing parameter which is often too coarse compared
to the smallest features.
Global sizing parameter chosen model by model would allow a greater
reduction of the number of irregular vertices, typically a factor from 5 to 10.
The median and average running times for the whole pipeline are respectively 18
and 155 seconds, using a single-thread on a laptop 2.8 GHz CPU.
If we normalize by the number of vertices, our complete pipeline takes around
one second per thousand vertices in the output, in average.

\subsection{Discussion}
\label{ss:discussion}

\begin{figure}
    \begin{center}
        \includegraphics[width=1.\textwidth]{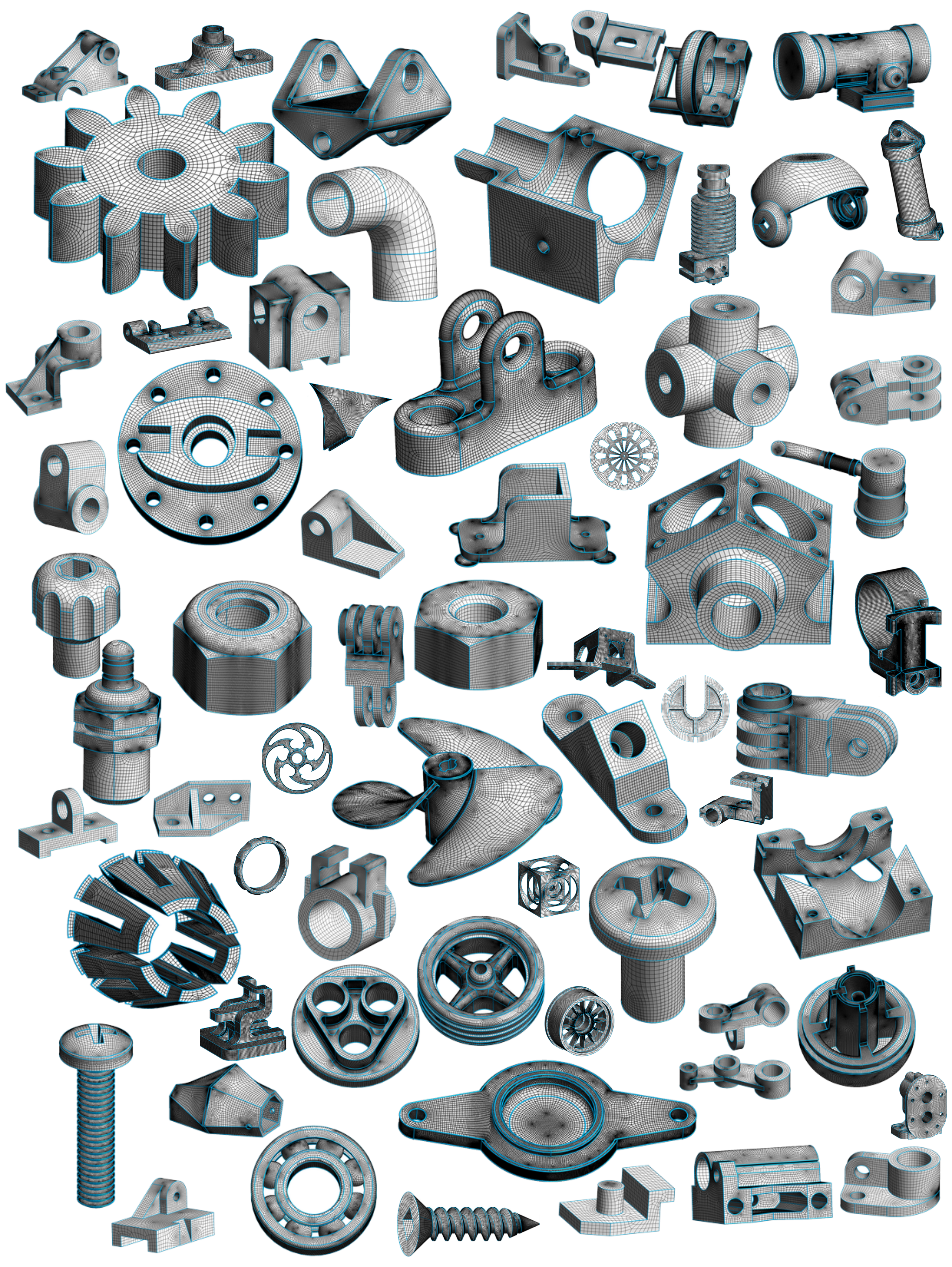}
        \caption{Quadrilateral meshes of models in the MAMBO \cite{mambo} and ABC
            \cite{mambo} datasets. 
        }
    \label{fig:mosaic}
    \end{center}
\end{figure}

Statistical results (\rsec{ss:database}) are interesting to confirm the robustness
of our approach but they are not very relevant to evaluate the mesher's behavior.
They are strongly affected by the target meshing size, \ie smaller size
would improve both regularity and quality. In the same spirit, midpoint subdivision
steps would artificially improve the statistics whereas the topological structure is
not changed. To better understand the mesher strengths and weaknesses,
it is necessary to visually inspect the quadrilateral meshes.
A detailed quad mesh of a complex model is illustrated on \rfig{fig:block} 
and some of the database results are shown in the 
mosaic \footnote{generated with the Mosaic generator (\url{github.com/qnzhou/Mosaic})}
in \rfig{fig:mosaic}.

\paragraph{Overview}
Our end-to-end meshing pipeline is able to robustly generate
quadrilateral meshes with high-quality elements on almost all CAD models, using
a single global sizing parameter. There are a few failure cases,
but they are due to either a \emph{very} wrong target size choice or a
CAD-adapted size map leading to too many vertices ($>5$M), for which we have
performance issues. 
When using an appropriate target size, we are able to eliminate 
many irregular vertices (\eg factor five to ten) in the initial unstructured
mesh, while preserving a relevant global structure thanks to the cross-field
singularities. In its current state, this new mesher is a viable alternative
to triangular meshing in Gmsh for generic CAD models and we expect further improvements
in the future. The flexibility offered by the unstructured approach makes it easy
to integrate user-prescribed constraints, such as specific element sizes at
given points or a custom size map given as a scalar field.


\paragraph{Comparison with unstructured techniques}
There are two large steps in our approach: the robust generation of the
initial unstructured quad mesh and its improvement via cavity remeshing.  On
the unstructured part, our technique is essentially an incremental improvement
of standard indirect meshing \cite{blossom,delquad}. Thanks to the accurate
cross field computation and its inherent size map (conformal scaling), we
generate better vertex positions which follow integral lines of the
\emph{locally integrable} scaled cross field (\eg \rfig{fig:gparam_vs_qqs}.2.b).
Concerning the topological improvement, we employ two topological techniques.
For local and constrained configurations, we pick the best rearrangement by
checking in an exhaustive list of disk quadrangulations (\rsec{sec:disk_quad}),
which is a new technique to our knowledge. For large cavities and simple CAD
faces, we use pattern-based remeshing similar to \cite{verma2015} but with
better cavities as we are using the cross-field singularities to control their
spawning and growth.

\paragraph{Comparison with global parametrization techniques}
When the size map is dominated by the cross field scaling and not by the
adaptation to small features, our approach produces meshes tending to the ones
produced by global parametrization techniques (\eg \rfig{fig:gparam_vs_qqs}),
except that there may be a few additional irregular vertices due to the non
optimal quantization of the curves. Most global parametrization literature is
from the computer graphics community and focus on smooth models without or with
few feature curves, without much consideration for CAD entities. Thus, it is
difficult to know if they could be adapted to strictly respect them. As they
are using a global sizing parameter without adaptation to small features, the
basic way to match them is to introduce strong distortions in the
parametrization and it is not clear if it is possible to extract geometrically
valid quads afterwards. Our approach has the advantage
of working with custom size maps containing strong gradients whereas global
parametrization techniques would first require to generate a cross field
matching the size map, which is an open problem.

\paragraph{Performance}
The initial unstructured quad mesh generation, starting from the CAD, 
is fast and is a few times the cost of triangular meshing. 
The computational cost of the topological improvement is less
predictable as it depends on the number and size of the cavities that
are candidate for remeshing. In general, the cost is similar to the initial
step but it may be much higher, especially for large CAD faces with lot
of vertices (>100k). The vast majority of the time is spent doing smoothing, 
which is essential to guarantee geometric validity of the elements.
To ensure limited running times, it is possible
for the user to limit the time spent on each CAD face. In the future, 
we plan to develop better cavity growing strategies to deal more efficiently
with large meshes by reducing the number of cavity candidates.

Thanks to the bottom-up approach, meshing and remeshing operations can be
applied in parallel on each CAD face, as the boundary meshes are fixed. The scaling
with the number of cores is excellent, and we are only limited by the RAM
as we are using shared memory in Gmsh parallelism (OpenMP).


\section{Perspectives and future work}
\label{sec:future}

In this work, we have presented a quadrilateral meshing pipeline which is
viable for industrial use. The robustness is ensured by starting from 
a basic unstructured quad mesh (quad-dominant split into quads) and
iteratively improving it while always checking the validity of each operation.
Additional improvements stemming from recent research on cross fields 
are incorporated in the pipeline as guiding information, but never at the cost
of robustness. Starting from this solid basis, we envision various lines 
of research to further enhance the topological structure of the final quad meshes.

\paragraph{Remaining non-essential irregular vertices}
When the target size is fine and the size map is dominated by the cross field
and not the CAD features, we would ideally expect to match the cross field
singularities. With our approach, it requires an optimal quantization of curves,
which is hardly achieved.  We see two solutions to this problem, either a
better initial quantization or a posteriori remeshing across CAD curves.

For the initial quantization, we should consider more the topological constraints
in quadrilateral mesh. In particular, the curves of the polygonal CAD faces whose topology
matches the patterns in \rfig{fig:patterns} should have a number of edges such that
the quantization is meshable according to the ILP problem. For the CAD faces without
cross field singularities, we can assign a direction $+x,-x,+y,-y$ to
each boundary curve and ensure that the sum of the edges over both axis is zero.
By combining these linear constraints in a global mixed-integer problem, we should be
able to produce a better initial quantization of the whole model.

A more pragmatic alternative is to eliminate the remaining irregular vertices
by modifying the mesh topology globally. The remeshing cavities could grow
across the CAD curves to surround opposite pairs of $+1/-1$ irregular vertices on
distinct faces. The difficulty is in ensuring the CAD curves are still preserved
after remeshing and in dealing with non-manifold face junctions. Another possibility
is to move the pairs of $+1/-1$ in the global mesh by successively applying local
mesh modifications.

\paragraph{Performance and large models}
The main limitation of our mesher is that it is slow when dealing with
large CAD face meshes ($>100k$ vertices) because there are too many
remeshing cavities which are tried, and each one involves smoothing
the geometry. We believe there is a lot of room for improvement
in the remeshing strategy (\rsec{ss:strategy}), which is quite
simple for the moment. A possibility could be to replace the current
isotropic growth of cavities by more surgical ones which would target
the non-essential irregular vertices and avoid the regions of the mesh
which are already regular.

\paragraph{From quasi-structured to block-structured}
For many applications, the best quadrilateral mesh is block-structured,
meaning that its base-complex is coarse and composed of a minimum of
large quadrilateral patches.
This kind of meshes is interesting because it can be easily converted 
into a good high-order quad mesh which is ideal for accurate high-order
methods. Even at low orders, the block structure allows for fast
computational optimizations (\eg implicit indexing, multigrid).
Our quasi-structured meshes are not ready for such applications because their
base-complexes are composed of too many patches.
Even when the number of irregular vertices is low (\eg matching the cross field
singularities), the issue is that the irregular vertices are not aligned
and consequently the topological chords tend to wind up over the whole model.
In the future, we wish to explore ways to align the irregular vertices, such as
\cite{bommes2011}, and build coarse layouts on complex CAD models. This is
definitively a challenging problem as the CAD features must be fully preserved
during the process.

\paragraph{Extension to hexahedral meshes}

Hexahedral meshing is a very difficult problem, especially
when trying to respect the features of CAD models. The current
robust techniques are either the subdivision of a tetrahedral mesh into
hexahedra or the octree-based approach \cite{marechal2009}. Both produce
meshes whose topological structure is not very good: there
is a large number of irregular edges and vertices, and the hexahedra
are not aligned with the model boundary.  We think that exploring ways to
extend the cavity remeshing technique to the 3D case could be a fruitful line
of research as it would allow iterative and robust topological improvements
to reduce the number of irregular edges and vertices in such meshes.

\section*{Acknowledgments}
This research is supported by the European Research Council (project HEXTREME,
ERC-2015-AdG-694020) and by the Fonds de la Recherche Scientifique de Belgique (F.R.S.-FNRS).

\bibliography{references}%

\clearpage




\end{document}